\documentclass[useAMS,usegraphicx]{mn2e}

\title[X-ray spectra of U Scorpii]
{Thomson Scattering and Collisional Ionization in the X-rays
 Grating Spectra of the Recurrent Nova U Scorpii}
\author[M. Orio et al.]
{M. Orio,$^{1,2}$ E. Behar,$^3$ J. Gallagher,$^2$ A. Bianchini,$^4$  
E.Chiosi,$^1$, G.J.M. Luna,$^5$
\newauthor
 T. Nelson,$^6$ T. Rauch,$^7$ B.E. Schaefer$^8$ and B. Tofflemire$^2$\\
$^1$INAF--Osservatorio di Padova, vicolo dell' Osservatorio 5,
   I-35122 Padova, Italy \\
 $^2$ Department of Astronomy, University of Wisconsin, 475 N. Charter Str., Madison WI 53704\\
$^3$Department of Physics, Technion, Haifa, Israel\\
$^4$ Astronomy Department, Padova University, vicolo dell' Osservatorio 3,
    I-35122 Padova, Italy \\
$^5$ Instituto de 
Astronomía y Fisica del Espacio (IAFE/Conicet), CC67, Suc. 28 C1428ZAA CABA,
 Argentina\\
$^6$ School of Physics and Astronomy, University of Minnesota, 116 Church St SE,
 Minneapolis, MN 55455 \\
$^7$ Institute for Astronomy \& Astrophysics, Kepler Center for Astro
\& Particle Physics, \\
 Eberhard Karls University, Sand 1, 72076 T\"ubingen, Germany \\
$^8$ Physics and Astronomy, Louisiana State University, Baton Rouge,
 LA 70803}
\begin{document}

\date{Received:}

\pagerange{\pageref{firstpage}--\pageref{lastpage}} \pubyear{2002}

\maketitle

\label{firstpage}

\begin{abstract}
We present a {\sl Chandra} observation of the recurrent nova U Scorpii,
 done with the HRC-S detector
 and the LETG grating  on day 18 after the observed visual maximum
of 2010, and compare it with
 {\sl XMM-Newton} observations obtained in days 23 and 35  after maximum.
The total absorbed flux was in the range 
 2.2-2.6 $\times 10^{-11}$ erg cm$^{-2}$ s$^{-1}$, corresponding  to  unabsorbed luminosity
 7-8.5 $\times$ 10$^{36} \times$(d/12 kpc)$^2$ for N(H)=2-2.7 $\times$ 10$^{21}$ cm$^{-2}$.
On day 18, 70\% of the soft X-tray flux was in a continuum typical of a very hot 
white dwarf (WD) atmosphere, which accounted for about 80\% of the flux
 on days 23 and 35.  In addition all spectra display very broad emission lines,
due to higher ionization stages at later times.
  With {\sl Chandra} we observed  apparent P Cygni profiles. 
We find that these peculiar profiles are not due  to blue shifted absorption and red shifted  
emission in photoionized ejecta, like the optical P Cyg of novae,  but 
they are rather a superposition of  WD atmospheric absorption
 features reflected by the already discovered Thomson scattering corona, 
 and emission lines due to collisional ionization in condensations
 in the ejecta. 
On days  23 and 35 the absorption components were
 no longer measurable, having lost 
the initial large blue shift that displaced them from the core of the broad emission lines.
We interpret this as indication that 
mass loss ceased between day 18 and day 23.
On day 35, the emission lines spectrum became very complex, with several
 different components.  Model atmospheres indicate that the WD
atmospheric temperature was about 730,000 K on day 18 
 and reached 900,000 K--one million K on day 35. This peak temperature 
 is consistent with a  WD mass of at least 1.3 M$_\odot$.

\end{abstract}


\section{Introduction}

The recurrent nova (RN) U Scorpii is one of a handful of such objects
known in the Galaxy, although this class may be much more numerous than
 we have been able to assess so far. ``Recurrent'' means
that the thermonuclear flash is repeated on a human timescale.
 The outburst supposedly recurs also in classical novae,
however the time scales can be as long as several times 10$^5$ years on
 WD at the low end of the nova
 mass distribution  (0.6-0.7 M$_{\odot}$), accreting matter
 at a low rate of $\leq 10^{-11}$ M$_\odot$ year$^{-1}$
 (see Yaron et al. 2005 and references therein). The models predict that RN occur
either with moderately
high accretion rate $\dot m$ on WD of about 1 M${_\odot}$,  or at very high
 $\dot m$ (M$\geq$10$^{-8}$ M${_\odot}$ year$^{-1}$) on
 WD with M$\geq$ 1.2 M${_\odot}$. The
 velocity of the ejecta increases, and the duration of the outburst decreases,
 with the WD mass.  About half of all RNe are {\it symbiotic
 novae}, that is systems with a  giant secondary and a hot WD, with
orbital periods of hundred of days (as opposed to most classical
 novae, which have orbital periods of hours and main sequence companions).

  U Sco  has an orbital period of 1.23 days
and  belongs to the class of RNe with a near-main sequence or moderately
 evolved secondary
(Schaefer 1990, Schaefer \& Ringwald 1995). The secondary
 spectrum is not detected in optical spectra at quiescence, but at a late post-outburst
 phase Anupama \& Dewangan (2000) measured the spectral features of
 a K2 subgiant. In the previous outbursts the ejecta
 were extremely depleted in hydrogen (Barlow et al. 1981), so the secondary must have lost
 a significant fraction of its hydrogen rich envelope.
At quiescence, He II lines are dominant in optical spectra and the hydrogen
 lines are absent or weak, a fact that has been interpreted
 as accretion of He rich material from a H-depleted secondary
(e.g. Duerbeck et al. 1993). Although
 Maxwell et al. (2011) cast doubts on the high
 He abundance, the possibility of very enhanced  helium in U Sco is very interesting
 as it may result in a type Ia SN (SN Ia) without a hydrogen rich envelope,
which is consistent with the upper limits on hydrogen in the SNe Ia spectra.

\begin{figure}
\includegraphics[width=85mm]{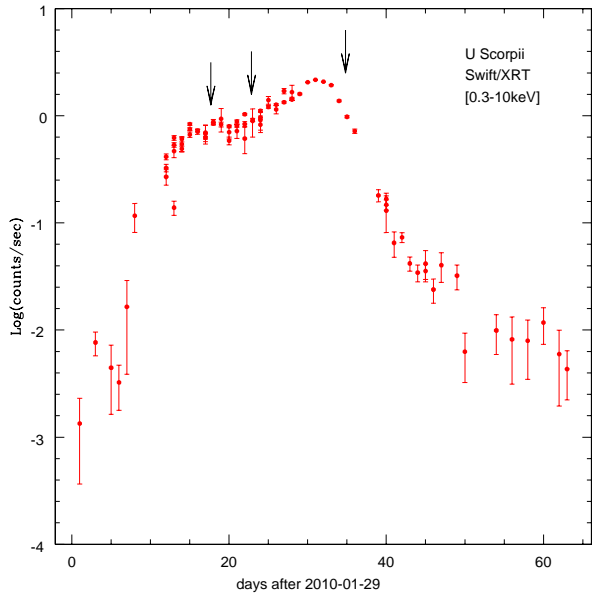}
 \caption{Swift XRT count rate versus time measured since since the recorded outburst on
 2010-01-29, in the 0.2-10 keV range.}
\end{figure}

 The short and luminous outburst of U Sco was observed 10 times (1863,
1906, 1917, 1936, 1945, 1969, 1979, 1987, 1999, 2010),
 with peak magnitude V$\simeq$7.5, decay times by 2 and 3 optical magnitudes,
 t$_2$=1.2 and t$_3$=2.6 days respectively, and  a return to quiescence
 within 67-68 days. These parameters indicate a very massive WD, such that
 sufficient pressure for a thermonuclear runaway is built up at the base
 of the envelope even with a small accreted envelope (e.g.
few $\leq$10$^{-7}$ M$_\odot$, see Starrfield et al. 1988).
We adopt 12$\pm$2 kpc as distance to the nova, measured
 thanks to its eclipse in the optical (Schaefer et al. 2010).

 Since U Sco has shown outbursts on an almost regular timescale of
 10$\pm$2 years (assuming a couple of eruptions were missed because
they occurred when
 the nova was behind the sun), Schaefer (2010) was able to predict the
 last eruption of January of 2010 with an uncertainty of only one year.
 When the event occurred, a strategy was in
place to observe the outburst at all wavelengths.

 In this paper we present
 grating X-ray spectra, the only means to probe the effective
 gravity, chemical composition and effective temperature of the underlying
 white dwarf once the photosphere shrinks back to the WD
 radius (e.g. Nelson et al. 2008 and references therein).
 One observation was proposed by us, 17 days after the visual
 maximum with the {\sl Chandra} LETG grating. { We adopt here the epoch of
 the observed visual maximum 2010-01-28.1 (JD 2,455,225,605) 
as the initial time (although Schaefer, 2010, extrapolated the light curve concluding
 that the peak had occurred a few hours earlier).}
Two {\sl XMM-Newton} observations were done
 on 2010-02-19 at 15:41:09 UT,
 and on  2010-03-14 at 14:33:32 UR, 22 and 35 days after maximum,
 respectively (note that for the different
 {\sl XMM-Newton} detectors the beginning
of the observations varies slightly, and
 for details  see also Orio et al. 2011,  Ness et al., 2011a and 2012).

 In addition to presenting and analysing the early {\sl Chandra}
 spectrum, in this paper we compare it with the {\sl XMM-Newton} RGS
 grating spectra. The proposers of one of these observations, Ness et al. (2012)
 have analysed the X-rays and UV data and examine the time variability in
 these two bands in detail. They find that a partial eclipse  of the
 X-ray continuum can be explained only if the 
 hot central source is observed after having
 been Thomson scattered, a conclusion that we reached independently 
(Orio et al. 2011), and with which we fully agree. These authors
 remark about the lack of absorption lines, expected and  usually observed in novae
 in outburst in X-rays, and attribute it to possible ``smearing''
 by Thomson scattering. They also describe and examine the spectra, proposing
 lines identifications and attributing the emission
 lines to resonant scattering. In this work,
 we found that the nova development with time, including the
 {\sl XMM-Newton} observations, is fundamental to fully understand the 
 physics of the X-ray grating spectra.
 Without the intent to replicate Ness et al.'s (2012) work, we 
 analyse and attempt to model with detailed
 physics all the grating X-ray spectra, thereby 
 reaching different conclusions than the above authors 
 on the origin of the emission lines observed with {\sl XMM-Newton} and on 
 the reason for the lack of absorption features at late stages.  

 In Section 2 we describe the observations,
 the data extraction method, and when the observations 
 occurred in the context of the long term X-ray evolution.
 In Section 3 we show the short term light curve, which we find relevant
 to understand the physics of the spectra. In
 Section 4 we discuss the apparent
 P Cyg profiles observed in the {\sl Chandra} spectrum.
 After having analysed the origin and meaning of these features,
 we performed atmospheric fitting for all spectra, and evaluated 
 temperature evolution and peak temperature (Section 4). 
 In Section 5 we discuss the physical mechanisms producing the emission lines
 and how we attempted a global fit; finally conclusions are presented in Section 6.

\begin{figure}
\includegraphics[width=86mm]{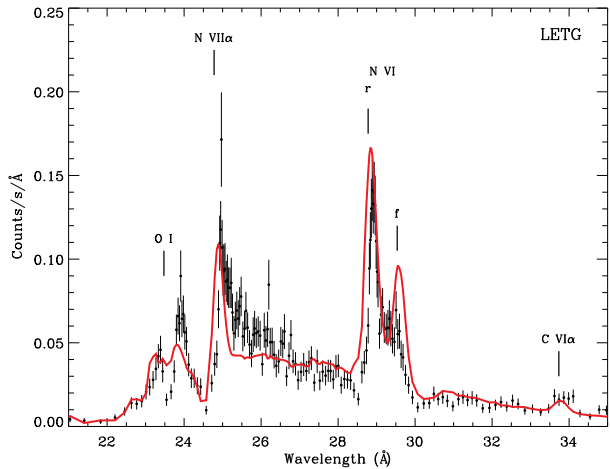}
 \caption{The U-Sco spectrum measured with the {\sl Chandra} LETG grating
 on 2010-02-14, day 18 post optical maximum in the V band (+1 and -1 orders
 have been summed). 
The red solid line shows the best fit with two components: an atmospheric model
 with a 740,000 K T$_{\rm eff}$ and 
 a collisionally ionized optically thin plasma with
 tenfold He and N abundance abundances with respect to solar, and plasma temperature
kT=93 eV (see Table 2).}
\end{figure}

\section{The observations, and their position in the 
long term light curve}

The {\sl Swift} satellite monitored U Sco almost daily on all dates
when it was technically possible.
We extracted the {\sl Swift data} and present the
 XRT X-ray light curve in   Fig. 1. Although this paper's  focus is on the longer
 observations done with {\sl Chandra} and the comparison with the RGS of {\sl XMM-Newton} data, 
the {\sl Swift}-XRT long term light curve puts the data in the context of the outburst evolution.
The arrows indicate the times at which the grating observations were done.
 The X-ray flux was still on the rise during both the 2010-02 observations
(days 18 and 23), while the maximum
 X-ray flux was recorded around the date of the third and final X-ray grating
 observation in March, on day 35. The X-ray spectrum remained
 remarkably soft during the whole post-outburst phase, with
 more than 95\% of the counts below 0.7 keV.
 A sharp decay followed the March observation until day 67, when the nova was almost
 back to X-ray minimum.  Further information on the UV and X-ray evolution
 observed with {\sl Swift} can be found in Ness et al. (2012).

\begin{figure*}
\includegraphics[width=120mm]{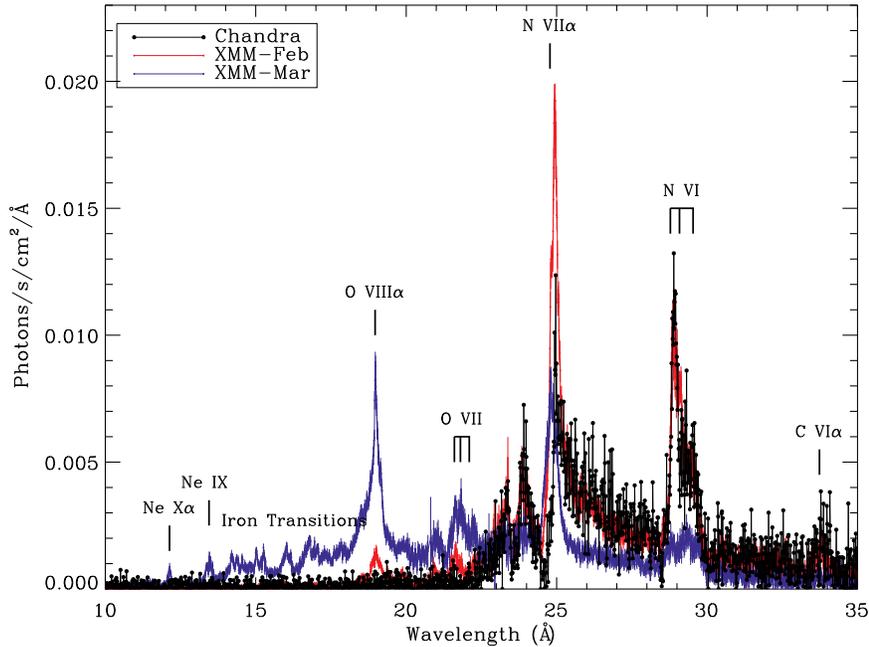} 
 \caption{Measured flux versus wavelength
 in the three grating spectra at days 18, 23 and 35 after the discovery,
The Chandra spectrum is displayed in black, the
 averaged RGS1-RGS2 spectrum of 2010-02-14 is plotted in
 in red and the  2010-03-02 RGS1-RGS2 spectrum is in blue.}
\end{figure*}

The first grating observation of U Sco was proposed by us and 
 was done  with the LETG/HRC-S instrument of the {\sl Chandra} observatory on 2010-02-14,
at 11:38:22 UT. As stated above, this was
 day 18 after optical maximum in the V band (see Schaefer et al. 2010). 
The exposures lasted for 23 ks (6.38 hours).
 The observed count rate spectrum is shown in Fig.2, while Fig. 3
 shows the fluxed spectrum, compared with the spectra measured
 with the {\sl XMM-Newton} RGS gratings on days 23 and 35.
Fig. 2 also shows a model fit, discussed in detail in Sections
4 and 5.  In order to extract the LETG spectrum,
 we applied the most updated calibration to the LETG pipeline processed
 level 2 {\sl event} and {\sl pha} Chandra data files.
 Response matrices and ancillary
 response (effective area) files were created using the tools mkgrmf
 and mkgarf, part of the CIAO v4.2 package.
We measured a summed $\pm$1 orders count rate of 
0.477$\pm$0.008 cts s$^{-1}$ in the 0.2-0.8 keV
 energy range in which flux is measurable above
 the background  (corresponding to approximately 15-60 \AA \ in wavelength range).
 We measured an unabsorbed flux 2.2$\pm$0.2 erg cm$^{-2}$ s$^{-1}$. 
In Fig. 3,  the strongest lines of the three spectra are labelled (for additional lines
 identification, see Table 1).

 We see in Fig. 2 that for at least three lines (the Ne VI He $\gamma$ line 
with rest wavelength  23.77 \AA,
the N VII hydrogen $\alpha$ line,
 with rest wavelength 24.78 \AA,  and the resonance line of the
 N VI triplet at 28.78 \AA)
the emission portion appears red-shifted and the absorption
 wing blue-shifted. This is intriguingly similar to
 the P Cyg profiles observed in optical spectra.

 We compared the {\sl Chandra}
 spectrum  with archival {\sl XMM-Newton} grating spectra obtained
 on 2010-02-19 at 15:41:09 UT
 and on 2010-03-3 at 14:33:32 UT,  (5 and 17 days later,
 respectively  days 23 and 35).  The {\sl XMM-Newton spectra} 
were extracted with XMM-SAS version 11.0.2. There
 was variability during both {\sl XMM-Newton}
 observations, as we discuss in the next Section,
 but in February (day 23 after the outburst) we measured average
 count rates 0.881$\pm$0.004 cts s$^{-1}$
and 0.824$\pm$0.004 cts s$^{-1}$ with the RGS1 and RGS2 gratings respectively,
while in March (day 35)
 the average count rates were 1.037$\pm$0.005 cts s$^{-1}$
and 0.951$\pm$0.004 cts s$^{-1}$. The unabsorbed
 flux seems was rising moderately, and we measured
 2.5$\pm$0.2 $\times 10^{-11}$ and 2.6$\pm$0.2 $\times 10^{-11}$
in the second and third observation, respectively. All {\sl XMM-Newton} instruments were
 observing the nova simultaneously, including the Optical
Monitor (OM), and the broad band X-ray imagers EPIC-pn and MOS1 and MOS2. 
The RGS spectra are shown in
 Figures 3 to 5. Figures 4 and 5 present the
 count rate spectrum (like Fig. 2 for {\sl Chandra}).
 The fit with atmospheric and plasma
 emission models, also shown in the Figures, are discussed in detail 
 in Sections 4 and 5.

\begin{figure*}
\includegraphics[width=85mm]{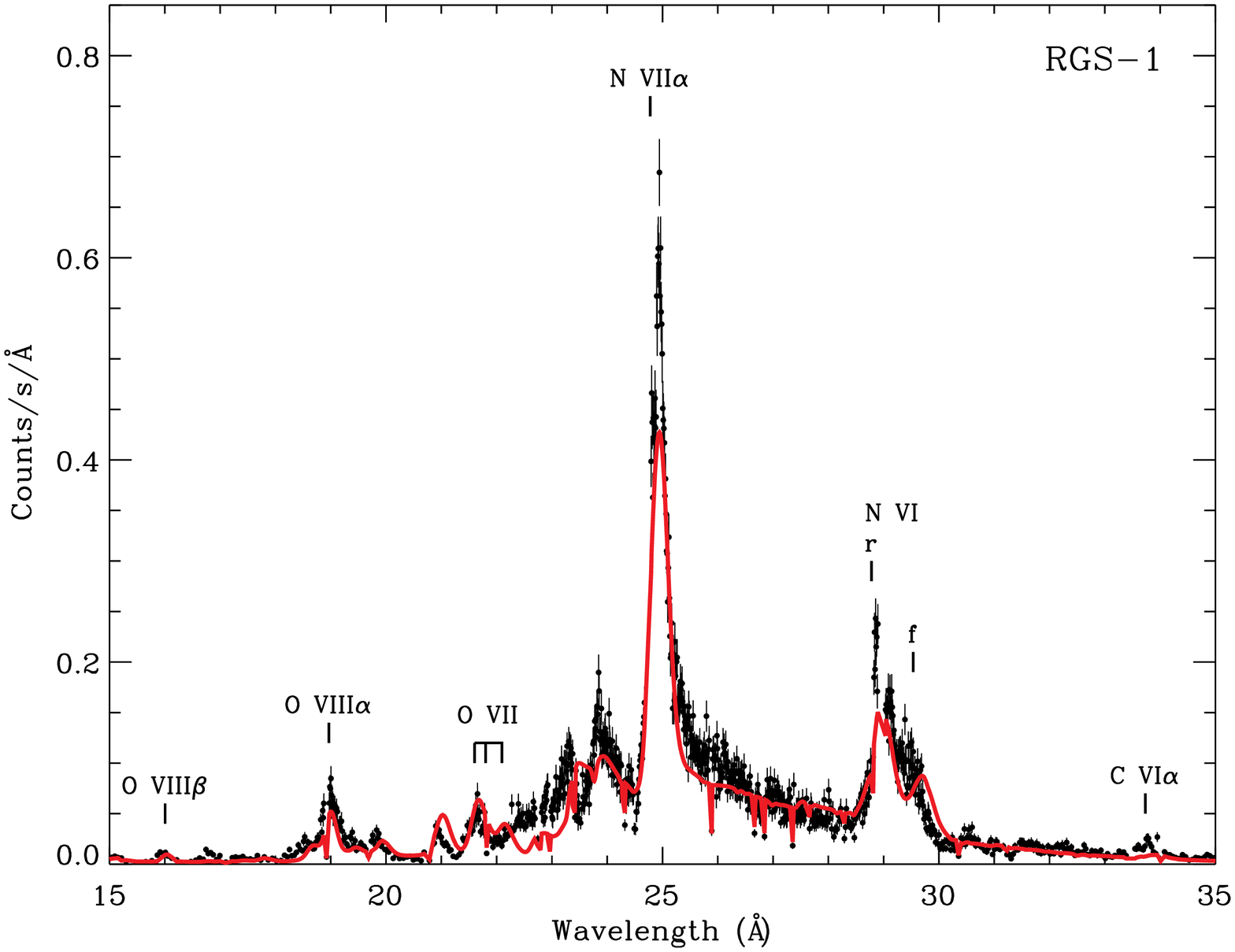}
\hspace{0.5cm}
\includegraphics[width=85mm]{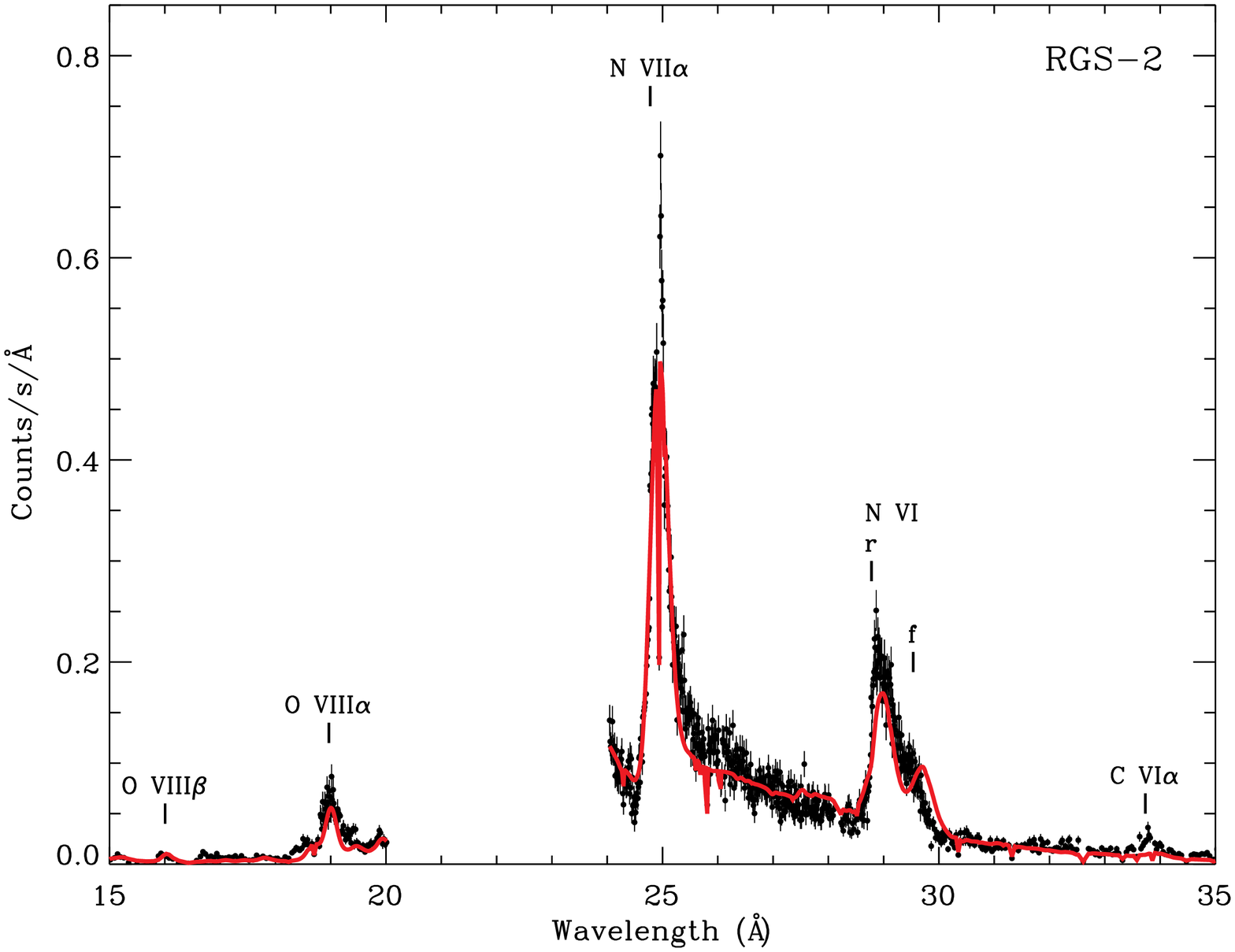}
 \caption{The {\sl XMM-Newton}-RGS-1 (left) and RGS-2 (right)
 spectra of U Sco on day 23  after the optical maximum. The blue
line shows the fit with
 a WD atmosphere with absorption lines blue-shifted by 2000 km s$^{-1}$,
 and only the {\it first} BVAPEC component in Table 2 (above)
 or adding the second BVAPEC components for the RGS-2 hard portion of the 
 spectrum (small panel below).}
 \end{figure*}
\section{The short-term light curve: a simultaneous X-ray and optical eclipse}

At optical wavelengths,
the eclipse of the accretion disk by the K dwarf companion
 was discovered and measured at quiescence  by Schaefer \& Ringwald
 (1995), and it was observed again in the outburst by Schaefer et al. (2011),
 indicating that the accretion disk had already been reestablished by the  third 
 week after the optical maximum.
 The orbital period
 of U Sco is 1.23054695 days (Schaefer \& Ringwald 1995).
  For the ephemeris we assume in this paper the result of the linear fit for the
whole 1999-2010 inter-eruption interval,  with a HJD time of the
 optical minima  2451234.5387 + N$\times$1.23054695.
 We caution that the
many observed eclipses during the last four 
eruptions show significant deviations of the minimum from this ephemeris, with
the offsets changing throughout the eclipse.  This means that the center
of optical light curve changes throughout the eruptions, and is not exactly at the same
position during the quiescent periods.

\begin{figure*}
\includegraphics[width=85mm]{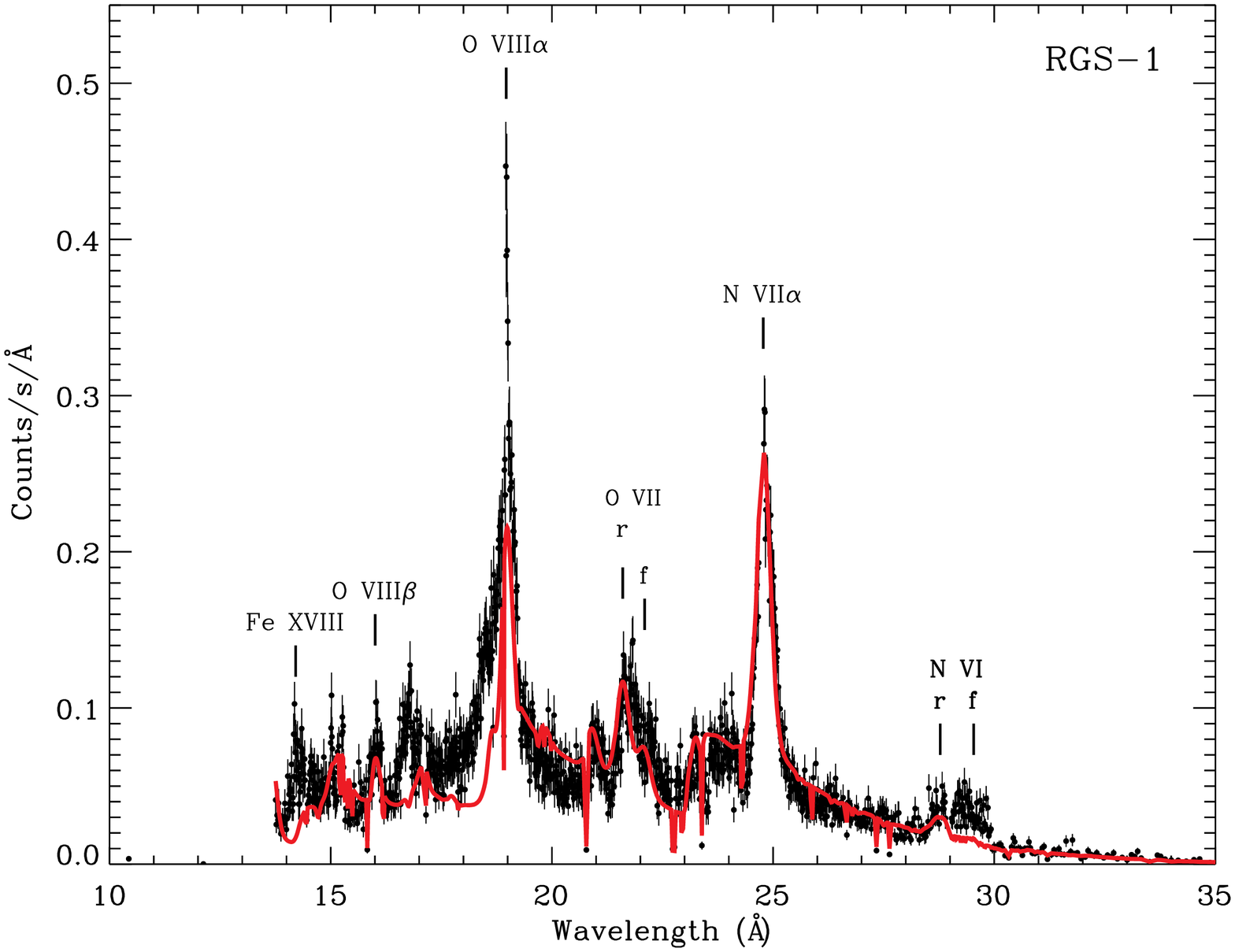}
\hspace{0.5cm}
\includegraphics[width=85mm]{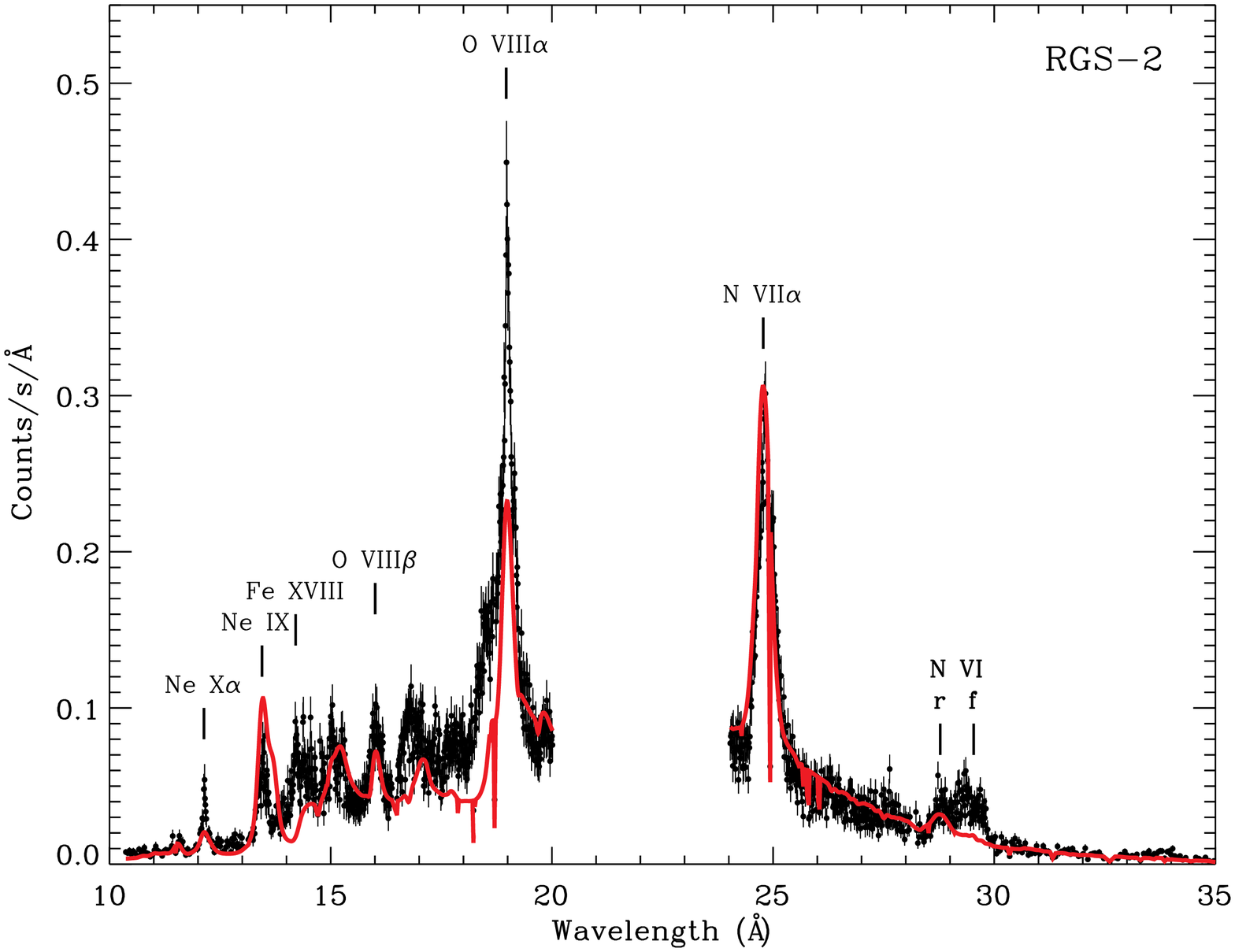}
 \caption{The {\sl XMM-Newton}-RGS-1 (left) and RGS-2 (right)
 spectra of U Sco on day 35 after the optical maximum. The
fit is shown in red, the parameters are given in Table 2.}
\end{figure*}

 We did not detect significant variability in the {\sl Chandra} zero order
light curve, but all the observation occurred
 outside the eclipse observed in optical. The exposure
started at orbital phase 0.4463 and covered
 almost a quarter (22\%) of the period.

\begin{table*}
 \centering
 \begin{minipage}{140mm}
  \caption{Rest, measured wavelength, and 
 fluxes in photons cm$^{-2}$ ks$^{-1}$ for the emission lines identified and measured
 in the three grating spectra. See the text for an evaluation of the errors
 in the flux estimate.}
\begin{tabular}{@{}llrrrlrlr@{}}
\hline
 &   &  Chandra 02-14 & & XMM 02-19 & & XMM 03-05 & \\
line               &  $\lambda_0$ (\AA)     & $\lambda_m$ (\AA) & flux &
$\lambda_m$ (\AA)  & flux & $\lambda_m$ (\AA)  & flux \\
\hline
C VI Ly$\alpha$  & 33.7342 & 33.78$\pm$0.02  & 0.39 &  33.78$\pm$0.05 & 0.43 & 33.75$\pm$0.05 &  \\
N VI f         & 29.5346 & 29.50$\pm$0.02  & 0.93 &  29.50$\pm$0.03 & 1.23 & 29.60$\pm$0.05 & 0.35 \\
N VI i          & 29.0819 &  29.16$\pm$0.02 & 0.91 & 29.20$\pm$0.02  & 1.65  & 29.22$\pm$0.05 & 0.42 \\
N VI r          & 28.7800 & 28.93$\pm$0.02  & 2.20 & 28.82$\pm$0.05  & 2.10  & 28.80$\pm$0.05 & 0.31  \\
C VI Ly$\gamma$ & 26.990  & 26.80$\pm$0.02  &   &       &  & \\
N VI He $\beta$ & 24.90   &                 &   &       & $\leq$3.8  & 24.95$\pm$0.01 & $\leq$1.3 \\
NVII Ly$\alpha$ & 24.78     & 24.94$\pm$0.03  & 1.71 & 24.92$\pm$0.03  & 5.64 & 24.80$\pm$0.02 & 2.51 \\
N VI He$\gamma$  & 23.771 & 23.81$\pm$0.05 & 0.43 & 23.81$\pm$0.05 & 0.60 & 23.82$\pm$0.02 & 2.70 \\
 O VII f & 22.097 &     & &  22.108$\pm$0.050 &                           & 22.13$\pm$0.05 & 0.30 \\
 O VII i & 21.801 &     & &  21.807$\pm$0.020 & $\leq$0.24                & 21.80$\pm$0.02 & 0.49 \\
 O VII r & 21.602 &     & &  21.603$\pm$0.050 & 0.35                      & 21.61$\pm$0.02 & 0.39  \\
 O VIII Ly$\alpha$ & 18.969 &  &              & 19.05$\pm$0.03        & 0.74 & 19.00$\pm$0.02 & 3.21 \\
 O VII  He$\beta$ & 18.627 & &  & 18.57$\pm$0.05 & & 18.55$\pm$0.02 &  \\
 Fe XVII & 16.78 &  & & 16.78$\pm$0.05 &   & 16.80$\pm$0.05 &   \\
 O VII Ly $\beta$  & 16.00 & & & 15.95$\pm$0.03 &  & 16.00$\pm$0.02 & 0.44  \\
 Fe XVII &  15.262 &  & & 15.30$\pm$0.05 & & 15.25$\pm$0.05 & $\geq$0.10 \\
 Fe XVII & 15.01 &    & & 15.15$\pm$0.02  &  & 15.03$\pm$0.05 & 0.12    \\
 Fe XVIII & 14.207 &  & &                 &  & 14.22$\pm$0.03 & 0.13  \\
 Fe XIX  & 14.667  & & &  14.70$\pm$0.05 & & 14.67$\pm$0.03 &  \\
 Ne IX f & 13.6984 & & & & &  &  \\
 Ne IX i & 13.5503 & & & &                                                  & 13.57$\pm$0.05 & 0.12 \\
 Ne IX r & 13.4474 & & & 13.45$\pm$0.02  &                                  & 13.48$\pm$0.05 & 0.15 \\
 Ne X $\alpha$ & 12.1321  &  &  & &                                         & 12.14$\pm$0.05 & 0.15 \\
 Ne IX He $\beta$r&  11.55 & &  & & & 11.60$\pm$0.10   &  \\
 Mg XI He r  &       9.09  & &  & 9.20$\pm$0.03 &  & 9.20$\pm$0.03 &  \\
\hline
\end{tabular}
\end{minipage}
\end{table*}

 Both XMM-Newton observations
 covered about 60\% of the orbital period and thus
overlapped with the optical eclipse time. The day 23 observation
 started at phase 0.84, and the one of day 35 started at phase 0.55.

In Fig. 6 we present the {\sl XMM-Newton} EPIC pn light curves 
 and compare them with the optical light curves of Schaefer et al. (2010)
 measured  in the same two weeks during which the two X-ray data sets were taken.
  All light curves are folded over the optically measured
 orbital phase and repeated twice. For the X-ray light curve, we plotted
a quantity that is 2.5 times the logarithm of the
 X-ray flux normalized to its minimum vale, for a direct comparison with optical magnitudes.
 The EPIC-pn exposures  are 63185 and 62318 seconds long respectively.  

On day 23 in the X-ray light curve the most evident  
oscillation is a pronounced one with a
 semi-regular (albeit not regular) period of about 3 hours,
 extensively discussed by 
 Ness et al. (2012). These authors also extracted
 the spectrum in and out of the oscillation, showing no variation
 of the  emission lines.
 The authors attribute the phenomenon to large clumps,
 of absorbing material at
 a distance about 3.5 R$_\odot$ from the WD, intersecting
 the line of sight in the direction of the accretion stream, during
 the process of re-establishment of the disk. 
 We do not wish to further discuss these oscillations, although we  
 suggest that this semi-regular period (even if it is associated
 with ejection of parcels of material), may represent
the disk rotation period, which may be slightly variable while the disk is
 still being formed.

 A clear orbital modulation in the
 day 35 spectrum was noted by us and by Ness et al. (2012).
 These authors also stress that almost all emission lines
 observed in the spectra are not eclipsed, while the continuum is,
 and we agree with their analysis. {  However, we cast some doubts  
 regarding the  dimming in eclipse of only one specific line, the O VIII Ly$\alpha$
 (the above authors remark this is the only line that 
 seems to vary, even there is clearly no variation of the O VII triplet).
 Extracting the spectrum over the
 short duration of the eclipse we find a variation of this line only at the 
 2 $\sigma$ level, so we suggest that this
is not statistically significant}. In any case, the main purpose of  
 Fig. 6 is to show that the orbital modulation seems already clearly 
present in the {\sl XMM-Newton} EPIC-pn light curve of day 23, although it is 
 not as pronounced as on day 35. On day 35, 
 the average count rate during orbital phases 0 to 0.12 and 0.88 to 1  is
0.75 and  0.84 cts s$^{-1}$ with RGS1 and RGS2 respectively,
 while we measured
1.10 and 1.05 cts s$^{-1}$ with RGS1 and RGS2 respectively during the rest of the observation. We find that 
 the variation in count rate in and out
 of eclipse, outside the $\simeq$3 hours oscillations, is by 
 at least 25\% even on day 23.

\begin{figure*}
\includegraphics[width=45mm,angle=-90]{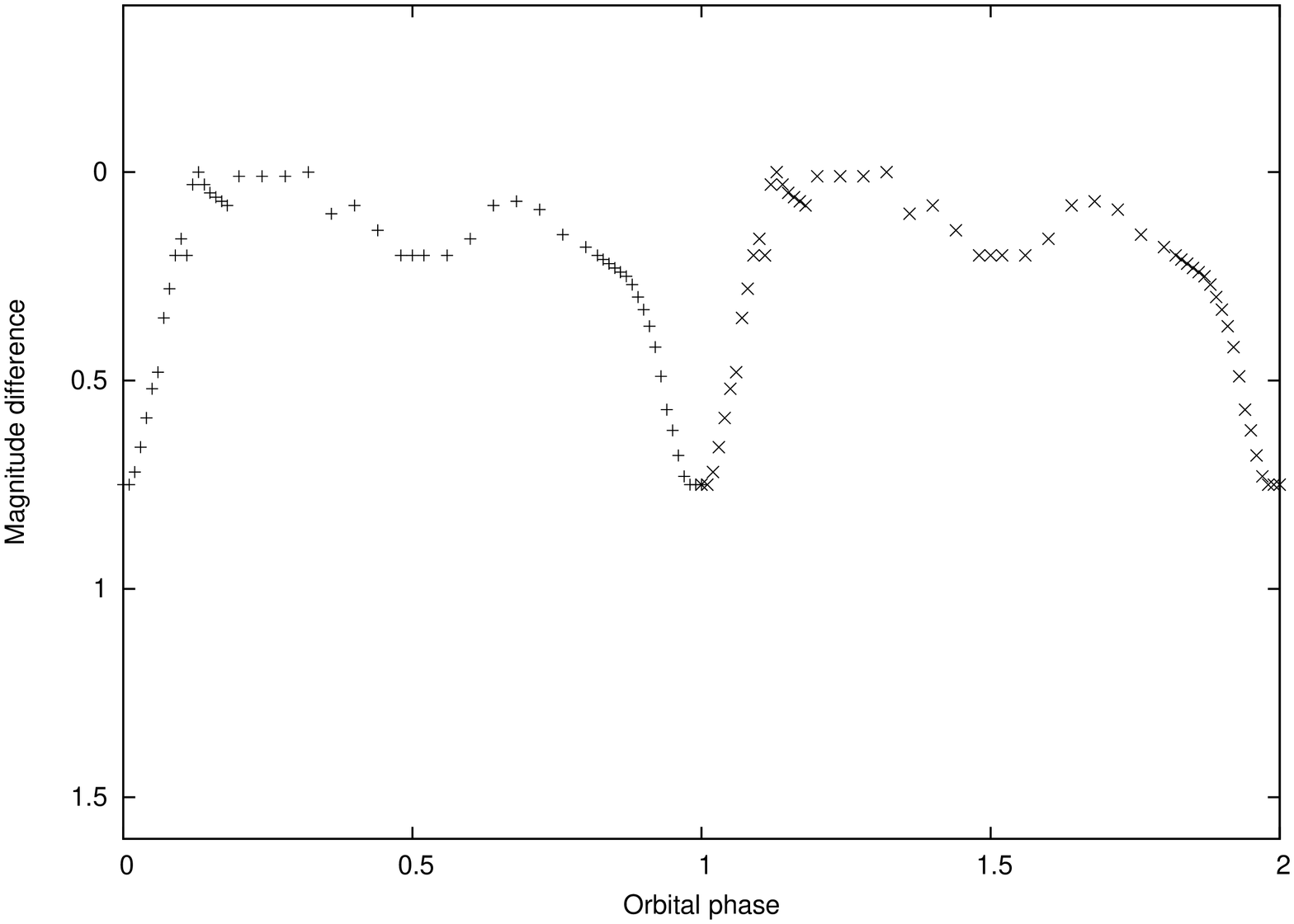}
\hspace{0.5cm}
\includegraphics[width=45mm,angle=-90]{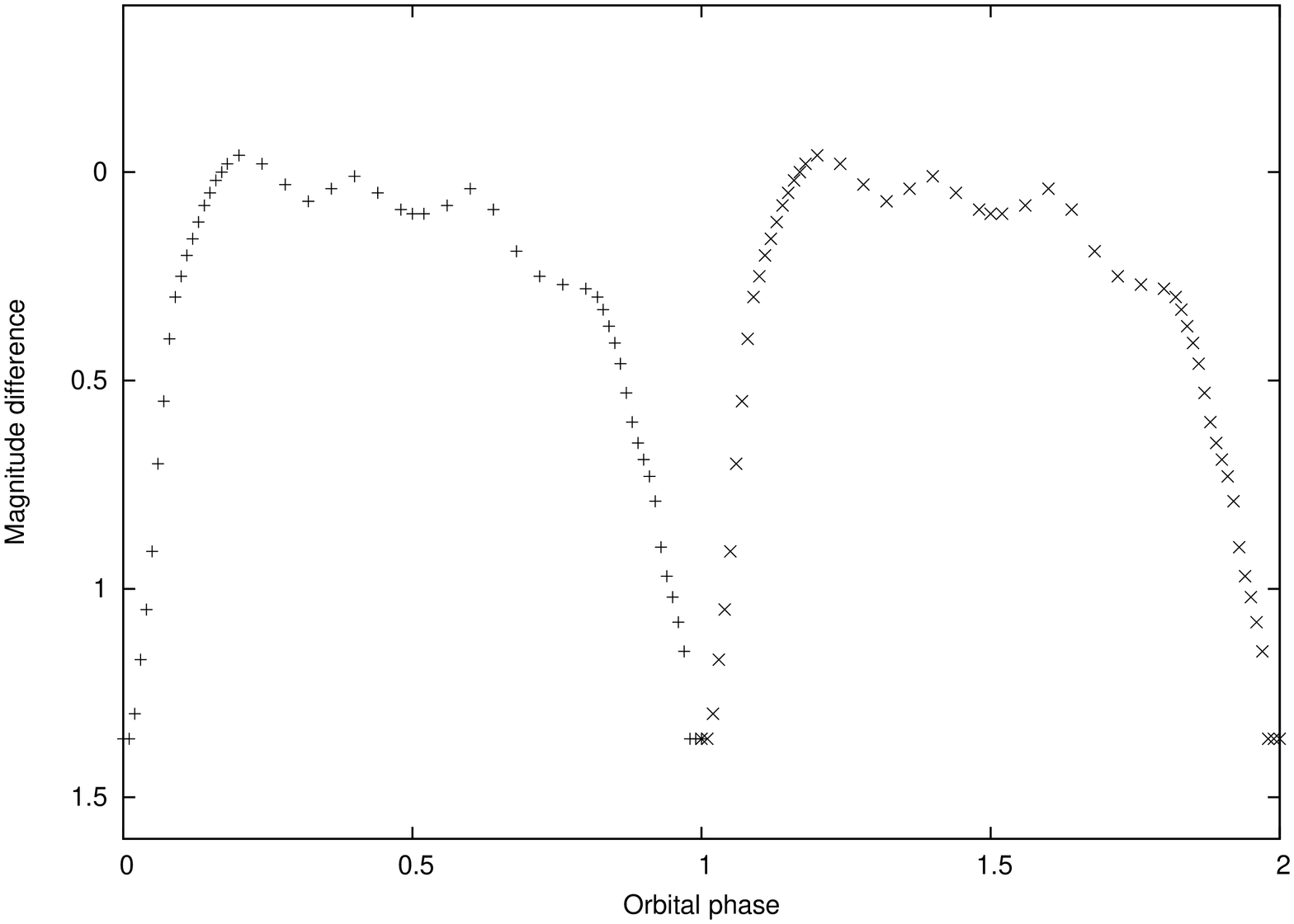}
\includegraphics[width=45mm,angle=-90]{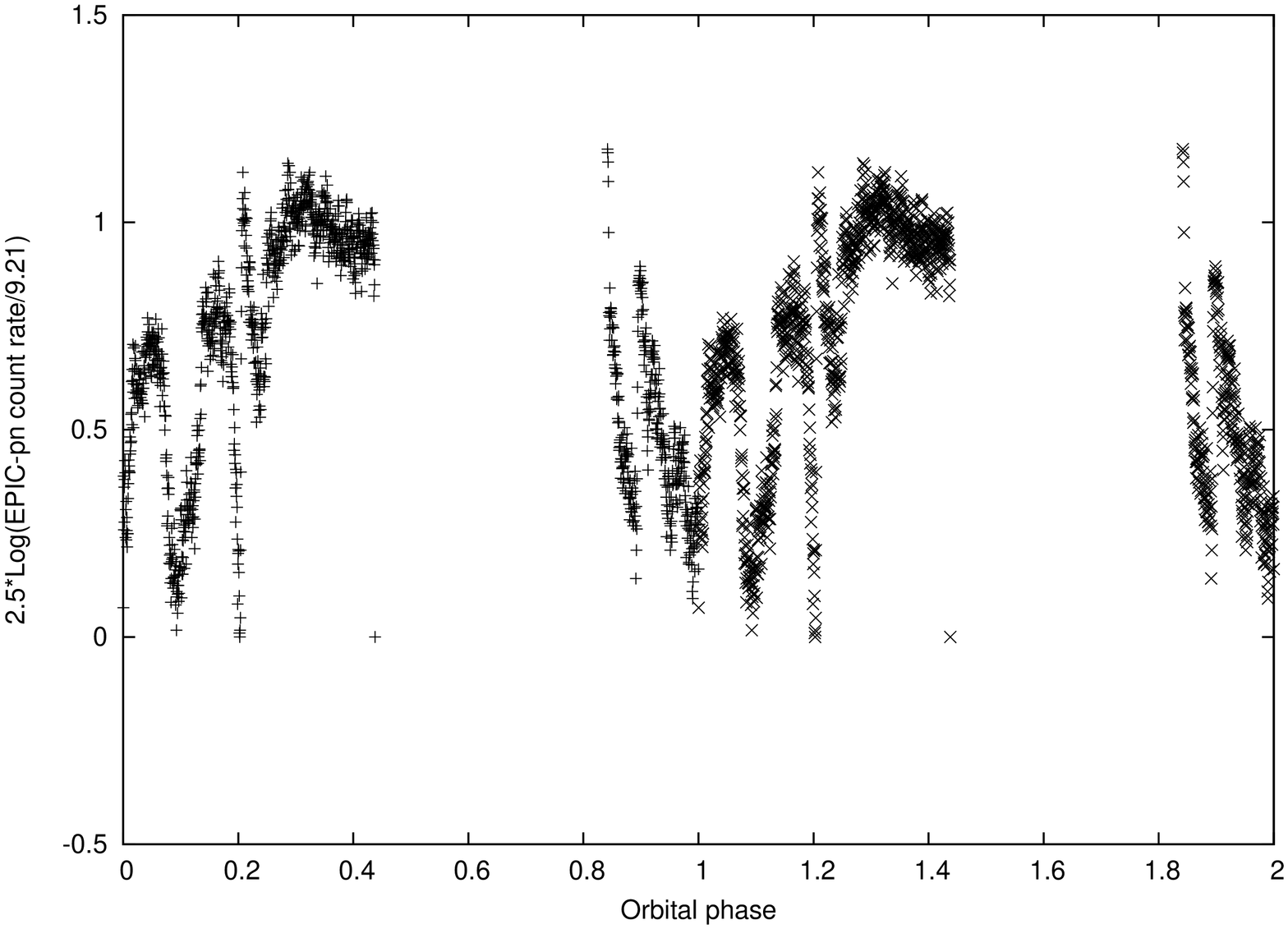}
\hspace{0.5cm}
\includegraphics[width=45mm,angle=-90]{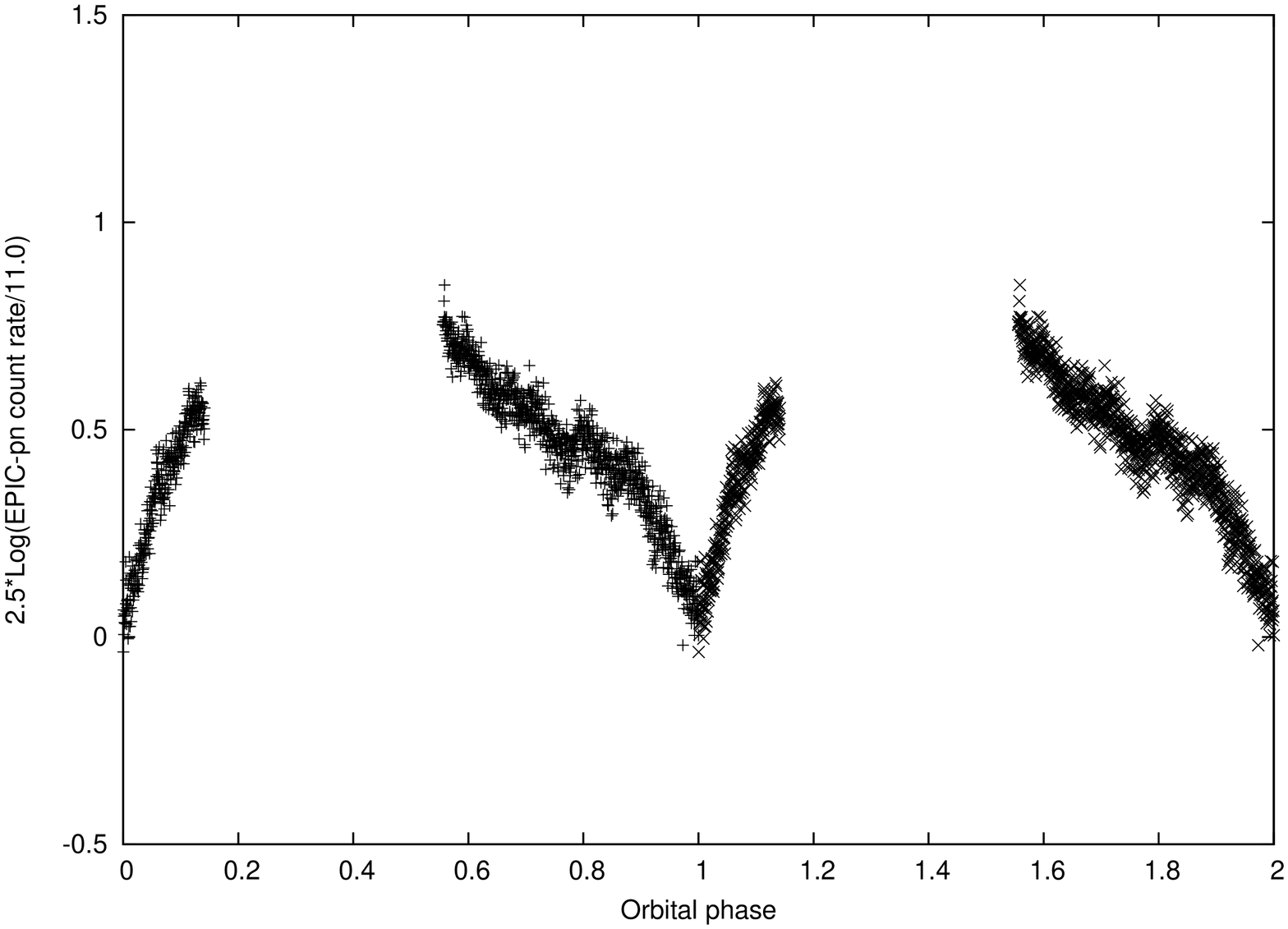}
 \caption{The average optical V magnitude lightcurve of U Sco during
 days 21-26  (left top panel) and days 32-41  after visual maximum 
  (right top panel; Schaefer et al. 2011), compared with the observed
 X-ray lightcurve measured with the EPIC-pn
 instrument aboard {\sl XMM-Newton} on days 23  (left bottom panel)
 and on day 35 (right bottom panel). The logarithm
 of the count rate normalized to the minimum value of the measurements
 and multiplied by 2.5 is plotted as a function
 of phase, to match the magnitude scale. The light curve was binned in time
 with 50 s bins before conversion  to phase, on day 23 (right) and 35
(left) after the outburst. 
}
\end{figure*}
 
  Because the eclipse takes out only about 50\% of
 the flux in the X-ray continuum, on day 35 a large portion of this continuum
 must originate in a much more extended region. We assume that this
is a Thomson scattering corona, re-processes
 a fraction of the WD radiation in an achromatic way (Ness et al. 2011a, 2012, Orio et al. 2011).
 The inclination angle in U Sco is 82.7$^o\pm 2.9^o$ (Thoroughgood et al. 2001), so
 it is quite conceivable that we did not observe most of the WD flux directly.
 The accretion disk was placed by Schaefer et al. (2011) at
 3.4 R$_\odot$ distance from the WD in a light curve model  for day 35.
 Because of the nearly edge-on inclination, {  the short-duration
 and deep X-ray eclipse does not appear to be due to the disk}. We assume that the source of
 the continuum  is eclipsed by the secondary, {  exactly like for the eclipse of the disk
 by the secondary observed at quiescence at optical wavelength}. 
 Following the reasoning of Schaefer et al. (2011, see their Fig. 13
 for the optical eclipse) we conclude that most
 of the continuum is observed as reflected at a distance 
 equal or larger than the a 3.4 R$_\odot$ distance from the WD  at which
 Schaefer et al. place the outer rim of the disk.

 We intend to show that the {\sl Chandra} spectrum with its absorption features
 is best explained assuming that this scattering corona was already present, and
 at comparable distance from the WD, on day 18, although no X-ray observations
 at eclipse minimum were done. 

 \section{Fitting the spectrum, step by step: the WD}

 \subsection{The absorption lines observed with {\sl Chandra}}

 Two puzzles are immediately apparent when we examine Figures 2 to 5. First of
 all, the {\sl Chandra} spectrum shows deep absorption features of
 nitrogen, while the {\sl
 XMM-Newton ones} do not display any features in absorption. Although 
 model atmospheres show that the absorption becomes
 less pronounced at high temperatures (Rauch et al. 2010), the models
 predict deep absorption features of nitrogen, always detected in
 the previous X-ray gratings' observations of other novae 
(e.g. Rauch et al. 2010, Ness et al. 2011b). The other surprising element, as we
 already noted,
 are the apparent P Cyg-like profiles in the {\sl Chandra} spectrum.
 P-Cyg profiles are typical of the optical spectrum of novae in X-rays, but they
 have never been observed in their X-ray grating spectra, except 
 (possibly) in the first {\sl XMM-Newton} spectrum
 of another fast recurrent nova, LMC 2009 (Orio 2012).  

 Emission lines attributed to the ejecta were observed in the X-rays spectra of several
 other novae (Drake et al. 2003, Ness et al. 2003, Nelson et al. 2008,
 Rauch et al. 2010). Are the absorption wings
of these ``P Cyg profiles'' in this U Sco spectrum produced in the nova shell,
 instead of being due to the WD?  At least one  profile cannot be
 a P Cyg in the conventional sense, and it is the one of  the N VI He $\gamma$ line.
 The absorption in this case is not the same line, but it is mainly of 
 another overlapping line at zero velocity, due to the OI 1{\it s}--2{\it p}  transition at
 $\sim$23.476 \AA, which does occurs in
 the ISM along the line of sight to U Sco.  The fact that this  absorption feature was
 hardly measurable in N LMC 2009, in one LETG spectrum that appears similar
 to this one (Orio et al. 2011, and 2012 article in preparation),
 confirms the identification because N LMC 2009 is in a direction of much lower
 interstellar absorption (N(H)$\simeq 4 \times 10^{20}$ cm$^{-2}$).
Thus, we realize that at least this particular P Cyg-like profile
 is in fact only a sort of  ``pseudo P Cyg'', composed
 of an absorption and an emission feature that do not have origin in the same medium.

 We examine now  
the  P Cyg profiles of the other two strong lines, shown in Fig. 7, to understand
 whether they may be due to the  ``real'', classic P Cyg phenomenon.
 As Fig. 7 shows in velocity space, the N VII emission
 appears blue-shifted by 2200$\pm$200  km s$^{-1}$ in absorption,
 and red-shifted by the same amount in emission.
The N VI resonance line with rest wavelength 28.78 \AA \
 is instead blue-shifted by 2700$\pm$300 km s$^{-1}$, in absorption,
and red-shifted by 1500$\pm$300  km s$^{-1}$ in emission, so the two components do not cross
 around zero, but rather towards the red. 
 In optical spectra  of novae,
crossing in the red  is observed in spectra taken shortly before the P Cyg
 profile disappears.  Typically, the P Cyg profile is no longer observed in 
the following weeks because the
 blue-shifted absorption becomes negligible compared to the emission.
 Fig. 8 shows the
 same line profile in velocity space in the RGS spectra of days 23 and 35.

 P Cyg profiles in  X-ray spectra in general are rarely observed, and the
 absorption wing, if measurable, is very shallow. This is because in X-rays we are likely
 to observe outflows and nebulae around compact objects, while in optical we
 observe them around stars with at least hundred of times larger radii.
 This includes the {\it optical}
 (as opposed to X-rays) spectra of classical and recurrent novae, where P Cyg profiles at optical
 wavelengths are observed when the photosphere around the WD has not receded to WD dimensions yet,
 but is still of the order of at least a solar radius. However, this typically occurs weeks
 or months before the supersoft X-ray phase, and we
 expect it only in the first few days, or even in the first hours, in a very fast nova like U Sco.
    P Cyg profiles at optical wavelengths are due mainly to
 resonance transitions. The blue-shifted absorption is observable
 only where the material moves in our direction,
 that is along the line of sight of the photo-ionizing central object.
The optical P Cyg absorption features occur, like the emission, in the ``wind'' or nebula around a 
 photo-ionizing central object, but they are 
 observed in a cylinder with the radius of the central object, and only
as long as the radius of the cylinder is not negligible 
 compared to the rest of the shell, from which we observe the emission lines.

 Luminous supersoft X-ray continuum - observed in the {\sl Chandra} spectrum as well
 as in the {\sl XMM-Newton} ones -
 in novae arises when the  central object has shrunk and regained compact dimensions,
 with a radius of the order of at most 10$^{9}$ cm,  or else the X-ray luminosity
would be largely super-Eddington.  
 Shortly
 after the outburst, velocities up to 5000 km s$^{-1}$ were measured
 by Yamanaka et al. (2010), and on day 35 (the day of the last
 {\sl XMM-Newton} observation), Mason et al. (2012) 
 estimated  an expansion velocity of 4000 km s$^{-1}$ from
 the full width at half maximum of the optical lines. Assuming this value
 for the expansion velocity,
 the nebular ejecta around U Sco at the time of the {\sl Chandra}
 observation (day 18) had reached a distance
 from the WD representing almost 300,000 times the WD radius of a very massive
 WD (1.3 M$_\odot$). 
 The region in which emission occurred
 was overwhelmingly large in comparison with the cylinder around the line of sight  towards 
 the WD. We conclude that the absorption features observed with 
{\sl Chandra} cannot have been produced in the ejecta.
 By day 18, any absorption features of the ejecta would already be 
 hidden and embedded in the broad emission lines.
There is  no evidence of nebular
 X-rays absorption lines in the ejecta, but in other novae  
 absorption lines were always observed, and they are
 usually attributed to the WD atmosphere. There is only a recent case in which
 the blue-shifted lines were attributed instead to collisional ionization
 in a thin outer shell, that of V2491 Cyg (Pinto et al. 2012), but the absorption
 lines of U Sco are so broad compared to this model, that we can rule out this origin.

\begin{figure}
\includegraphics[width=62mm,angle=-90]{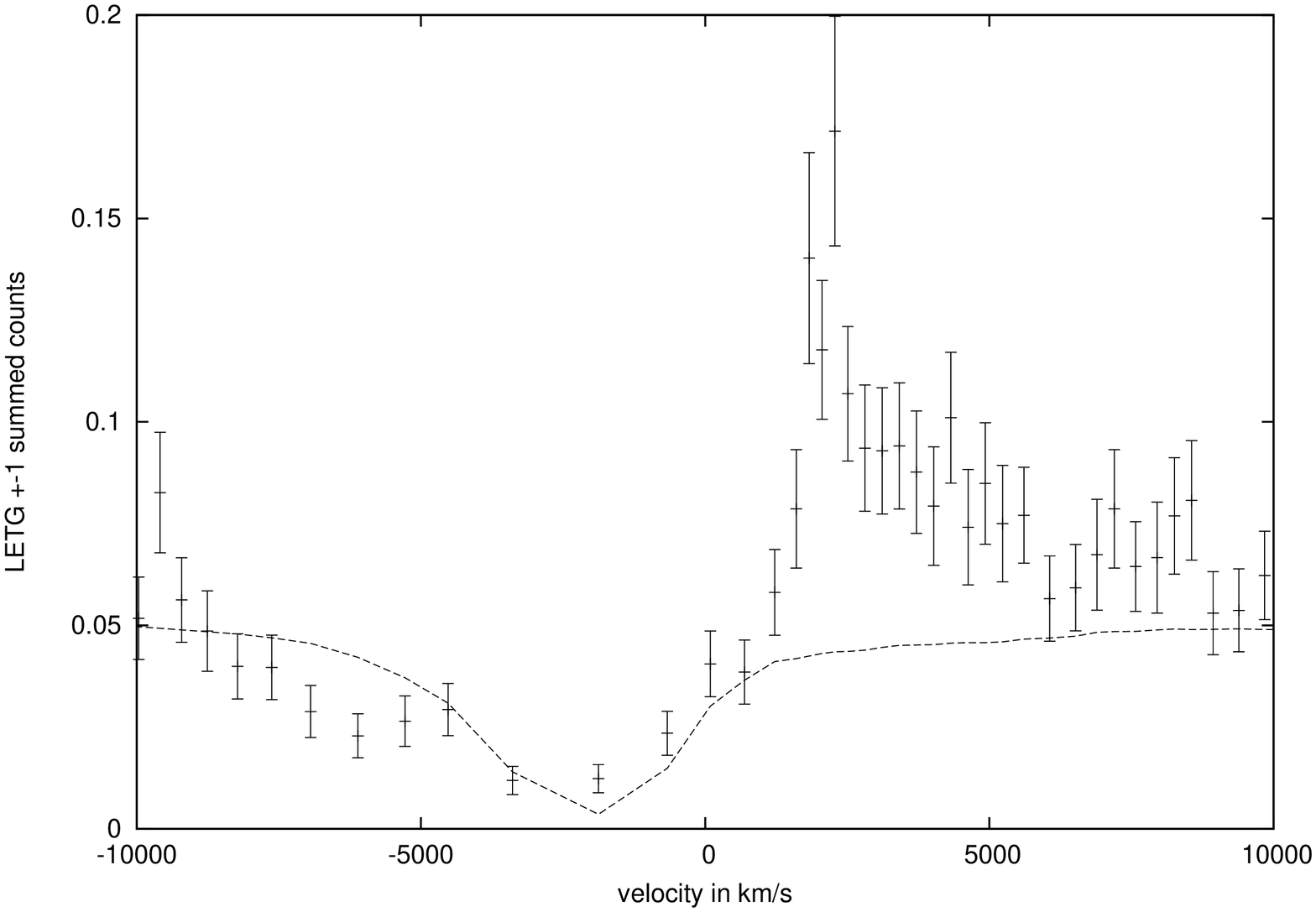}
\hspace{0.5cm}
\includegraphics[width=62mm,angle=-90]{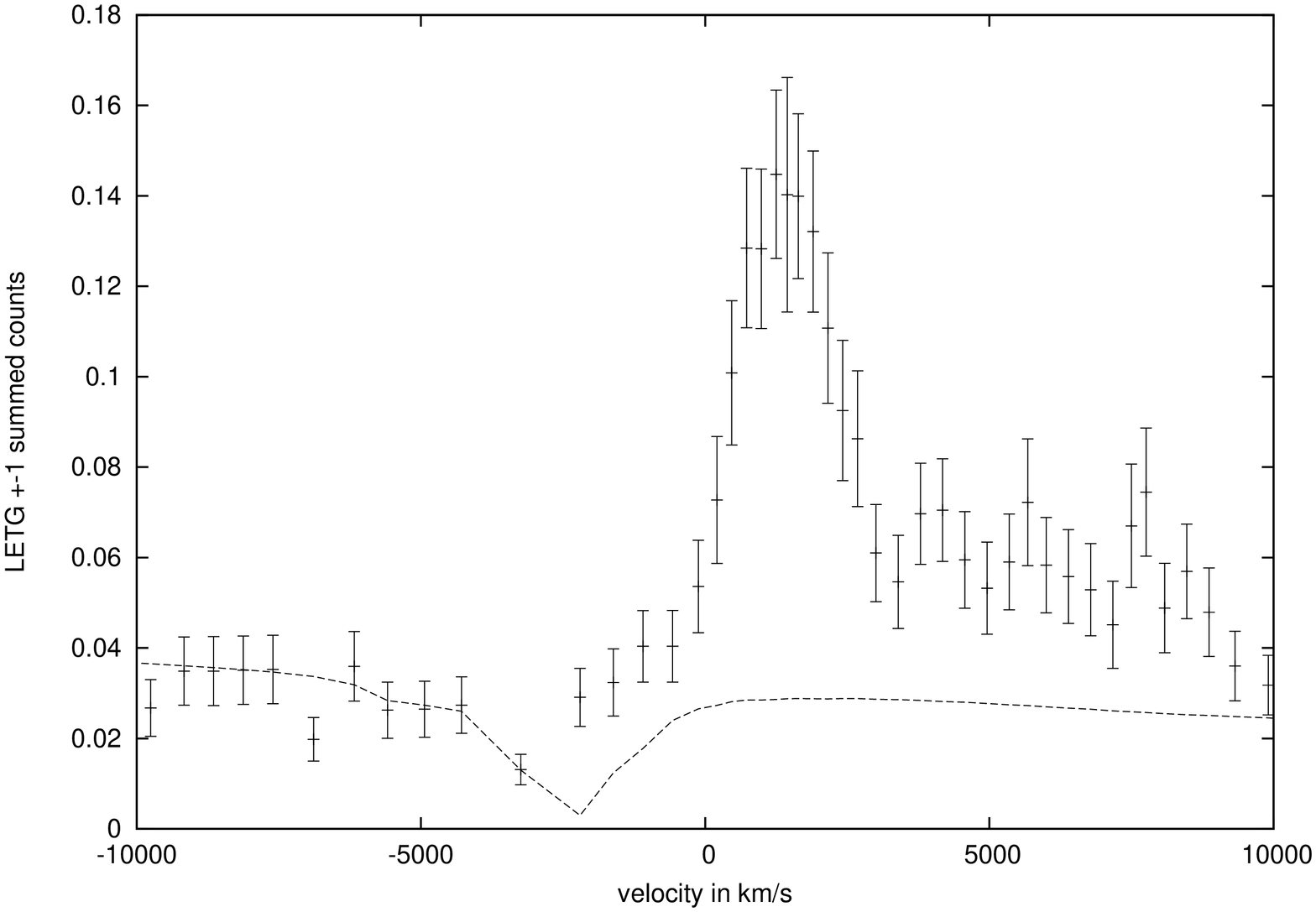}
\caption{The  P Cyg profiles of the Ne VII Hydrogen $\alpha$ line (above),
 and N VI resonance line (below), in the Chandra LETG spectrum ($\pm$1 orders summed)
 plotted in velocity space. The fit with
 an atmospheric model (see Table 2) with absorption features
 blue-shifted by 2000 km s$^{-1}$ is shown by the dotted
line. The two weaker lines of the triplet
 appear on the right of the resonance line of N VI in emission.
 There is a contribution of N VI He $\beta$
 line in emission at 24.889 \AA \ (see Table 1).}
\end{figure}

 We concluded that the velocity of the absorption features also for U Sco
 was due to a different phenomenon,  
 one causing an intrinsic velocity that had an almost abrupt
 end, so that the features were still significantly
blue shifted on day 18, but not any more a few days later.
This process must be mass loss from the nova. 
Absorption lines in X-ray grating spectra of other novae in the early phases
 were significantly blue-shifted,
 with comparable blue shift  to the apparent ``absorption wing'' of the ``P Cyg'' in the Chandra
 spectrum of U Sco (e.g.  Ness et al. 2003 and Rauch et al. 2010 for V4743 Sgr,
 Ness et al. 2011b for V2491 Cyg). In RS Oph 
 many of the same absorption
 lines as in the former two novae were observed  without measurable
 blue-shift (Nelson et al. 2008). Since the emission lines were strong in a very different range,
 there was no overlap with WD lines and no apparent P-Cyg profile. 
  We conclude that the P Cyg profiles in the {\sl Chandra} spectrum were 
 not ``real P Cyg'', but they were rather due to  superimposed absorption and emission 
 arising in different regions and different mechanisms.
 The absorption features  were produced  in the  outmost layer of the 
 WD atmosphere, which was still loosing mass, and reflected at a distance of 
a few solar radii by the {\it Thomson scattering corona} of the
 WD. The corona's ``reflection'' allowed the lines to be detectable, despite
 the high inclination angle of the WD in U Sco.
  Since this corona must have already been present on day 23,
 we assume that it existed already on day 18 (even if we could not measure 
 a light curve to prove it).

 An intrinsic blue-shift that lasted 
 while the WD was still loosing material in a wind, 
 explains why the features were still observable on day 18,
 even if the emission features are so broadened in U Sco that they should hide the
 atmospheric absorption. Once the wind from the WD ceased,
 the  absorption lines velocity became close to zero, 
 so they were embedded in the center of the emission lines and were not measurable any longer. 

 Whatever absorption features may have actually been produced 
  in the ejecta by photo-ionization, if there was such a phenomenon, 
  were embedded in the emission features long before day 18, due to the
 small dimensions of the absorption region compared to the fast  ejecta.
 It is also conceivable that the Thomson scattering corona changed in geometry and
 the WD absorption 
lines  may have become more smeared out between day 18 and 23, and  perhaps better 
 ``hidden''. In any case the transition
 giving rise to absorption lines in
 the atmosphere, even when not observable and measurable,  must have occurred until
 the WD turned off.

\subsection{Atmospheric models: the hottest WD}

 Because of the conclusions above, in order to fit the {\sl Chandra} spectrum, we 
 assumed that the observed absorption components were not produced
 by photo-ionization of the ejecta like in the
 optical P Cyg profiles of novae, but that they originated {\it inside} the WD atmosphere,
 and were reflected at large distance ($>$3 R$_\odot$) by the Thomson scattering.
 We fitted only the continuum
 of  the {\sl XMM-Newton} spectrum. We assume that on day 23 and 25 the atmospheric absorption lines 
 were simply unobservable, hidden 
 by the unusually broad emission lines of this nova once the mass outflow, and  with
 it the blue shift, ceased.
 Of course in the {\sl Chandra} spectrum the observed absorption may have 
 been more pronounced if it was not super-imposed on emission, so we can only
obtain lower limits on the abundances and  effective gravity. 

 Although the {\sl Chandra} spectrum absorption features 
 were eroded by the blue wing
 of the partially overlapping emission lines,
we note that they were still remarkably close in depth and broadening to those 
 observed in other novae without X-ray eclipses, such as
V4743 Sgr (Rauch et al, 2010) and RS Oph (Nelson et al. 2008). 
 We adopted the atmospheric models developed by coauthor Thomas Rauch, publicly available
on the web site
http://astro.uni-tuebingen.de/~rauch
and developed for very
hot white dwarfs, e.g. RS Oph (Nelson et al. 2008) and V4743 Sgr
(Rauch et al. 2010). We fitted the spectra with XSPEC, v12.6, 
 assuming that the absorption features were blue-shifted by 2000 km s$^{-1}$ (as
 in Rauch et al. 2010).
To date, Rauch's models represent the best approximation to the atmosphere
 of a hydrogen burning white dwarf, although
 residual outflows that produce blue-shifted absorption
 are not accounted for in these static models.
 Models including elements with atomic number
 up to the one of oxygen are available for a grid of values
 of effective gravity, with a log(g) grid from 5 to 9 with 1 step
 increments. The most recent grid of models, including all
 elements up to magnesium, was calculated for the hottest white dwarfs,
 with temperatures above 500,000 K and is suitable for the hot continuum of 
 the U Sco spectra. Since above 500,000 K, 
the WD must be very compact. and
 the luminosity would become largely super-Eddington with log(g)$<$9, 
 are published only for log(g)=9. 

\begin{figure}
\includegraphics[width=62mm,angle=-90]{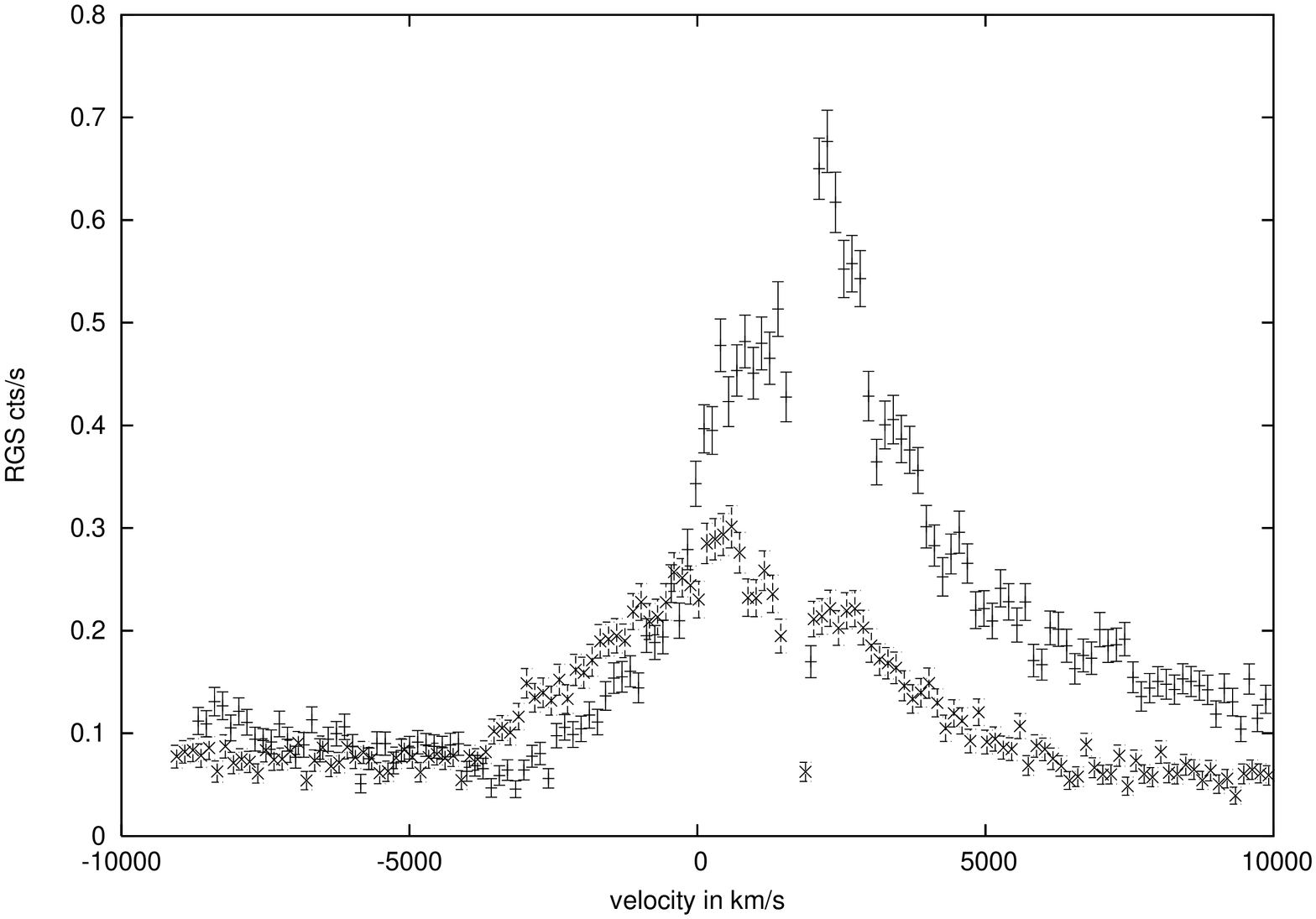}
\hspace{0.5cm}
\includegraphics[width=62mm,angle=-90]{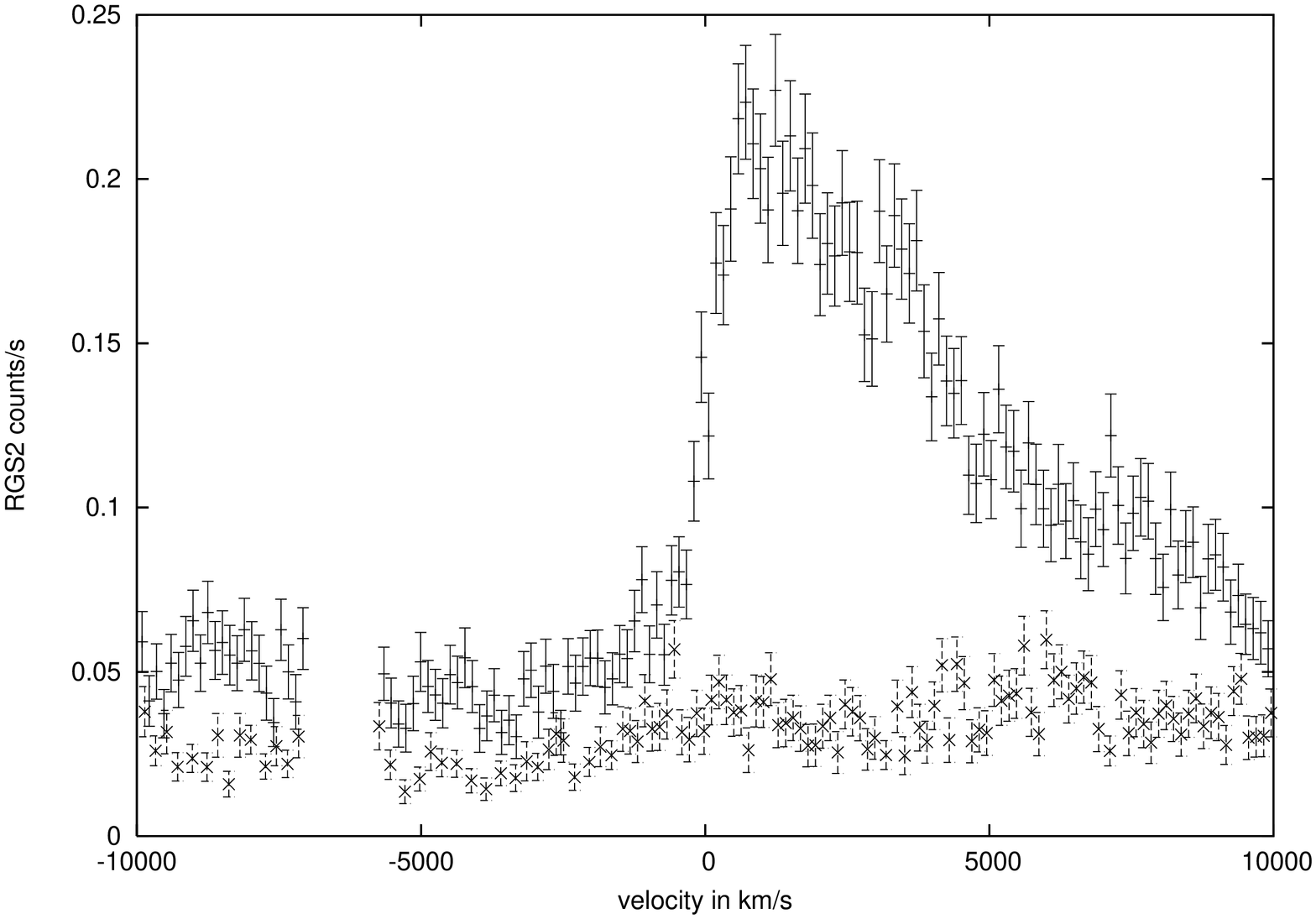}
 \caption{The NVII Lymann $\alpha$ line (above) and N VI resonance line (below)
 measured with the
 RGS2 grating in February (dots) and March (crosses), respectively,
 plotted in the velocity space.}
\end{figure}

 Model 201 in Rauch's web page was developed specifically
 with similar abundances to the ejecta of U Sco measured
 using the optical spectra in the previous outbursts: depleted
 hydrogen ([He]=0.489 or a helium mass fraction of 70\%, and almost
 50 times enhanced nitrogen with respect to the solar value
 ([N]=1.668). We fitted models of this grid with XSPEC, after blue-shifting the absorption features by
 2000 km s$^{-1}$. We do not obtain the best fit with model
201, but rather with model 003,
 with [He]=0.382 and [N]=1.803. This is not
 surprising, because the abundances in the burning layer can be
 quite different from those of the ejecta, where mixing with unburned material
 has occurred.

 The final product in our fit, shown in Fig. 2 for the {\sl Chandra} LETG,
 is a composite model of WD atmosphere and superimposed emission lines,
 but for the {\sl Chandra} spectrum we started by performing an experiment, 
ignoring the channels containing the strong emission lines (24.8--25.6 \AA \
for N VII Ly $\alpha$, and 28.7--30.0 \AA \ for the N VI triplet),
so that the model fit was better constrained by the continuum and by
the absorption features. For the composite fits  plotted in
 blue in Figures 2, 4 and 5 we
 binned the data with 30 to 50 counts per bin, but for the initial experiment,
  in order to retain the high spectral resolution of the LETG, we
binned the data only by a factor of 4 resulting in wavelength bins of
 0.01 \AA.  With a low resulting number of counts in each bin,
we used the C-statistic (Cash, 1979)
rather than the more commonly used $\chi^{2}$ statistic.
 We found that the C parameter is smallest for the smallest C/N ratio (typical of CNO ashes),
and steadily increases as the C/N ratio increases. The decrease of the ``C''
 parameter with C/N ratio is about 20\% from model 011 to model 003 of Rauch's grid.
All models with solar abundances yield poor fits to the {\sl Chandra} LETG
spectrum.
The temperature for the ``truncated-emission'' spectrum
 in the 2$\sigma$ confidence range ($\pm$ 2$\sigma$) is 650,000-750,000 K. The best fit yielded
 a temperature of 730,000 K, N(H)=2.5 $\times$ 10$^{21}$ cm$^{-2}$,
 and an absorbed flux 1.64 $\times$10$^{-11}$ erg cm$^{-2}$ s$^{-1}$,
 corresponding to an unabsorbed flux 4.2 $\times$10$^{-10}$ erg cm$^{-2}$ s$^{-1}$.
 The bolometric luminosity of the WD in this model is
 7.2 $\times$ 10$^{36} \times$(d/12 kpc)$^2$. 

  We fitted also the spectra observed later
 with {\sl XMM-Newton}, although there are no absorption
features to better constrain the temperature range. The fit returns almost the
 same temperature 
 for day 18 and 23.  In the  spectrum of day 35 the fit yields instead a  higher temperature,
in the range 900,000 to 1,100,000 K, and is unvaried  
 if we add the emission component, discussed in Section 5.
 The WD temperature at maximum X-ray luminosity
 is tightly correlated with  WD mass. We know from the models that
the effective
 temperature exceeds 900,000 K only for WD mass around 1.3 M$_\odot$
(Yaron et al. 2006).

The models yield an absorbed WD luminosity
 around 7$\times$ 10$^{36}$, which at 12 kpc distance is
 a factor of 5-10 smaller than the luminosity so far measured in other WD
 with even lower peak temperature (e.g. Balman et al. 1998, Orio et al. 2003a,
 Nelson et al.
 2008). For the other novae, the WD luminosity at peak was  several times
 10$^{37}$ erg s$^{-1}$. We explain also this puzzle, like the
 one posed by the apparent P Cyg profiles, if  
 we assume U Sco we did not observe the WD directly. The reprocessing factor
 for the WD radiation must have been of the order of 10\%. 

We note that after the previous outburst in 1999, Kahabka et al. (1999),
fitted the {\sl BeppoSAX} low resolution, low S/N LECS spectrum
 with a WD atmospheric model. They derived a luminosity in the range 2.7 to 18
 $\times$ 10$^{37}$ $\times($d$^2$/12 kpc) 19-20 days after the
 optical maximum, and an effective temperature of 75 eV (870,000 K).
 The count rate was however only
 0.06 cts s$^{-1}$, a value that appears too low
 by  a factor of about 2 if translated into both the {\sl Chandra}
 or {\sl XMM Newton}
 count rates with PIMMS for a blackbody at 75 eV and N(H)=2.5 $\times 10^{21}$ cm$^{-2}$. The
 {\sl BeppoSAX} observation lasted for about 14 hours, so the low count rate 
 is not explained with orbital variability. We cannot reconcile it with the
 lower luminosity of the present WD component. Probably the atmospheric
 model had significant differences from the most recent and detailed models
 of Rauch, but we also cannot rule out that the fit was correct, and the Thomson scattering
 corona may have been very different, or not present, in the
 previous outburst. 

\section{The origin of the emission lines: shocks, condensations
 and an evolving violent medium}
   
In the {\sl Chandra} spectrum 
20\% of the flux was in the emission features, and in the RGS spectra about 30 \%,
 while the remaining flux was due to the continuum.  Despite small uncertainties in
 the relative calibration of the {\sl Chandra}
 and {\sl XMM-Newton} instruments, as we see in Fig. 3, 
 the spectra are remarkably similar on days 18 and 23. However, two striking changes
 happened: the absorption
 features seem to have disappeared except for O I at 23.476 \AA, which
 we identified as interstellar in origin, and 
 oxygen emission lines suddenly appeared.
The O VIII line, which is not
 measurable in absorption in the WD atmosphere on day 18
 but is clearly detected in emission with {\sl XMM-Newton} on day 23,
 was almost symmetric around zero velocity since it appeared, consistently
 with no erosion by a nearby blue-shifted absorption line. This does not imply
 that the line did not exist at all in the WD atmosphere, because 
 the level of the WD continuum
 in that energy range is too low to measure absorption (see Rauch et al., 2010)
 
 In emission, the full width at half maximum of this line is approximately 4000 km s$^{-1}$,
 consistently with the  velocity measured in the optical spectra of
 Mason et al. (2012).
 As can be seen 
comparing Fig. 7 with Fig. 8, the emission features in the third spectrum appear symmetrical
 around the rest wavelength of the line, and have an almost triangular shape,
 indicating that lines
 are  produced in an optically thin expanding medium. In contrast, the
 profiles of the emission lines in the optical spectra of novae after
 the P Cyg profiles have disappeared usually  appear flat-topped, indicating that
 the expanding medium is optically thick.

We measured the emission line fluxes by fitting the strongest lines with  Gaussian profiles,
 and  we attempted a fit to a few weak
 lines with triangles.  The photon flux is
 given for each line in the three spectra in Table 1. We estimate
 that our measurements are accurate to within 10\% for isolated lines and flux
 above approximately 0.4 $\times$ 10$^{-4}$ photons cm$^{-2}$ ks$^{-1}$
(which were fitted with a Gaussian).
 The accuracy is probably not more than  $\simeq$25\% below this value (where 
 we fitted a few lines with a triangle, especially for the day 35 spectrum). 
 {  We did not attempt to measure line fluxes below 0.12 photons cm$^{-2}$ ks$^{-1}$, 
 because the uncertainty seems very large below
 this value, but in some cases we can still identify lines with probably almost an 
 order of magnitude lower flux (e.g.  Mg XI in the last spectrum).} 

 The emission lines, much less strong in this wavelength range in
 other novae, in the U Sco  spectra are very pronounced
and broad (compared with thermal and instrumental broadening, see
e.g. Brinkman et al. 2000,
 Ness et al. 2003). We attribute the line
 broadening to the expansion velocity of the nova shell, like the broadening often  
 observed in the optical spectra.
The recurrent novae LMC 2009 (Orio et al. 2011) and T Pyx (Tofflemire et al. 2012)
 have been the only other ones with broad emission features. 
We also note that the carbon lines are much weaker than those of nitrogen. Only
 the C VI Ly $\alpha$ line in emission at 33.90$\pm$0.12 \AA \ (rest wavelength
 33.7342 \AA) is clearly detected.

\subsection{A comprehensive spectral fit: days 18 and 23}

 In Fig. 2, 4 and 5 we plotted in red the fit to the lines and continuum
 obtained with physical models. These models can be regarded as
 a sort of first order approximation to the correct model,
 to explore the relevant physical mechanisms in these spectra.
 We have not reached a perfect fit, as the  value of the {  reduced} $\chi^2$ in these
 fits varies from 2.7 (Fig. 2) to 4 (Fig. 5). Rather than
 obtaining a statistically significant fit, we focused on 
 a good qualitative agreement and on reproducing the emission
 line ratios as well as possible. 
 We think that, as novae observations are continued in the next few years,
 we should be able to gather many more details of the nova
 physics, that will allow to fine
 tune these models of the very complex phenomena occurring
 in novae WD atmospheres and in their ejected shells.
 
 On days 18 and 23, the relative line strength, combined with
 the lack of a broad radiative recombination RRC feature  of N VI at
22.5 \AA, which should be
 clearly present in the case of photoionization (e.g. X-ray
 spectra of planetary nebulae or some AGN, see Fig. 5 of Kinkhabwala et al., 2002),
 indicate that the transitions are due to collisional ionization.
We used the XSPEC package, v12.6, adding to the 
 atmospheric model of the previous Section
 and a component of  collisional ionization, namely the BVAPEC model,
 that allows to include velocity broadening {  (1$\sigma$ value,
 which corresponds to the half width at half maximum of the lines)}.
 Table 2 gives relevant parameters of the spectral fits shown
 in the Figures. We assumed that helium in the plasma component 
 is enhanced 10 times with respect to solar values. 
 In order to improve the fit to the emission lines, we need a large overabundance of
 nitrogen not only in the also in the BVAPEC component, at least 10  times the solar value,
and even up to 70 times like in the fit we show in Fig. 2
for the {\sl Chandra} spectrum (this is the best fit with only two components,
although we obtained sightly worse, but
 similar fits with an enhancement by a factor of 10 and a slightly
 different set of parameters).
 We note that the WD unabsorbed luminosity resulting from these fits is reduced by up to 
 30\% with respect to the values contained in the ``experiment'' in Section 4.1,
 in which we fitted only the continuum, cutting the prominent emission
 features.
  
 Figures 2 and  4 show that we can reproduce the relative strength of
most lines by assuming collisional
 ionization lines with the Table 2 models for days 18 and 23. 
 However, the BVAPEC collisional ionization model in XSPEC assumes a low
 density limit and does not predict that
 the forbidden line is quenched compared to the intercombination line.
 We see in Figures 2 and 4 
that the N VI $\lambda$29.5446 forbidden line ii the N VI triplet
 on days 18 and 23, and the O VII $\lambda$22.097 line at day 23
 are less strong than in the models shown in
 the Figures. 

 We attribute this phenomenon to line formation
 in a zone of such high density that the collisional de-excitation rates are
close to the radiative decay rates.  When the two rates are comparable,
the radiative decay fraction from an excited level
are reduced and so the emission becomes weaker.
The electron density n$_{\rm e}$ at which the N VII forbidden line is de-excited is a few
 times 10$^9$ cm$^{-3}$ (Ness et al. 2001, Chen  et al. 2004).
 This density is not unheard of for novae shells in the first few weeks after
 the outburst. n$_{\rm e}$ of order of 10$^{12}$ cm$^{-3}$ was derived
 by Neff et al. (1978) from the analysis of the optical spectra of V1500 Cyg.

 This high density is  higher than the value derived from  the
 emission measure in the fit in Table 2, n$_{\rm e}$ = a few 10$^{7}$
 with the assumptions of the model, {\it if the ejected shell volume is uniformly
 filled}. 
 Since estimates of the ejecta mass of U Sco converge vary from
a few $10^{-6}$ M$_\odot$ to a few  $10^{-7}$ M$_\odot$ 
(see Schaefer 2011, Starrfield et al. 1988), assuming
 uniform filling of the shell, we expect that n$_{\rm e}$ does not exceed a few 
 times 10$^5$ at day 18, and it should be decreasing as the ejecta 
 expand. This discrepancy implies an emitting volume of the order of only 
 $10^{-4}$ the volume of the shell. The explanation seems to be
that line emission occurs in
dense condensations. We note that the large clumps that
explain the light curve $\simeq$3 hours period in the Ness et al.
 (2012) model occupy a much smaller fraction of the shell
 volume corresponding to only a few 10$^{-9}$. 
 Clumpiness is not completely unexpected, and it 
 has often been invoked to explain novae optical
 spectra, starting with the pioneering work of Gallagher \& Anderson (1976).
 Interestingly, already in 1992 Lloyd et al. discussed how the X-ray
 emission from a nova shell, albeit observed only
 with a broad band instrument ({\sl ROSAT}), implies compact clumps
 in the ejecta as emitting regions. 

 In our spectral fits, whose parameters
 are given in Table 2, the appearance of oxygen lines on day 23 depends on
 the interplay of plasma temperature and emission measure.
 Increasing the plasma temperature by only $\simeq$30\%, or increasing the emitting volume
 by a  factor of a few (assuming for instance that additional mass is being ejected)
 the oxygen lines are predicted, but then it becomes difficult to
 reproduce the other lines, correctly modeled
only with a plasma temperature in the 100-120 eV range.
 The strength of the O VIII Ly$\alpha$ feature is much stronger
 than all the lines of the O VII He-like triplet, indicating an overlapping
 component at higher temperature.
 In conclusion, in order to reproduce both the soft and hard portion of the spectrum, we need
at least two components at different temperatures.
 An additional component at 180 eV
 reproduces the relative strength of the oxygen features.
 The component that explains the nitrogen lines, at 120-130 eV, fails
instead to reproduce the ``harder'' lines.
 It is thus likely that on day 23 the condensations in which emission lines
 originated were not at uniform temperature.

\begin{table*}
 \centering
 \begin{minipage}{140mm}
\caption{Physical parameters of the fits shown in
Figures 2, 4 and 5 (model atmosphere + collisional ionization
model BVAPEC in XSPEC).
 The emission measure is derived from
 the BVAPEC fit normalization constant assuming a distance of 12 kpc.
 T$_{\rm WD}$ is the WD effective
 temperature in the atmospheric model. The
 flux is in the 0.2-1 keV range, and F$_{\rm WD}$ is the WD atmospheric flux.
 T$_{\rm p}$ is the plasma
 temperature in the BVAPEC model (two region at temperatures were assumed for a
 double BVAPEC fit for 2010-02-22). v is the 1 $\sigma$ velocity broadening, 
 F$_{\rm tot}$ is the total flux. The fit with two plasma temperatures
 also has two different average velocities in the two zone, two
 emission measures (with the cooler
 plasma contributing much more), and three
 values of N(H) (for WD, for the first plasma region at T$_{\rm p,1}$, and for
 the  second plasma region at T$_{\rm p,2}$). The emission lines were also redshifted 
 with redshift z  around few $\times$ 10$^{-3}$. A different value
 of N(H) was used for each of three components in the 02-22 spectrum (while N(H) was
 unique in the other fits), and we also indicate two velocities for this
 date, one value for each BVAPEC plasma component. {  Finally, N/H indicates the 
 nitrogen abundance, the ratio of this element over hydrogen (by mass) over the solar ratio}.}
  \begin{tabular}{@{}lllrlr@{}}
  \hline
Parameter & 2010-02-14 & 2010-02-22 & 2010-03-05 \\
\hline
N(H) (10$^{21}$ cm$^{-2}$) & 2 & 2.7/2.2/2 & 2.4 \\
T$_{\rm WD}$ (K)  & 739,000   & 727,000 &  1,050,000 \\
F$_{\rm WD}$ (erg cm$^{-2}$ s$^{-1} \times 10^{-11}$) & 1.79 & 1.33 & 1.77 \\
F$_{\rm WD}$ (unabs.) (erg cm$^{-2}$ s$^{-1} \times 10^{-10}$) & 3.94 & 3.9 & 3.6 \\
T$_{\rm p,1}$ (eV) & 93 & 130 & 223 \\
T$_{\rm p,2}$ (eV)      &     & 182 &     \\
Em. measure $\times$ 1.72 $\times$ 10$^{57}$ cm$^3$ & 12   & 32/3  & 20  \\
N/H         & 70   & 20  & 20 \\
v (1$\sigma$, in km  s$^{-1}$) & 1500 & 1680/1350 & 1800 \\
F$_{\rm tot}$ (erg cm$^{-2}$ s$^{-1} \times 10^{-11}$) & 2.40 & 2.46 & 2.57 \\
F$_{\rm tot}$ (unabs.) (erg cm$^{-2}$ s$^{-1} \times 10^{-10}$) & 4.9 & 4.9 & 4.22 \\
\hline
\end{tabular}
\end{minipage}
\end{table*}

\subsection{A complicated structure on day 35}

We are able to understand the emission line spectra of
 days 18 and 28 with collisional ionization alone, making the 
 assumptions of clumpiness, and
 adding a second temperature component on day 23, with only
 10\% of the emission measure of the cooler plasma, and larger oxygen abundance ([O/H]=1.56  versus
 only 0.2 in the cooler emitting region). This model thus would also imply 
 either inhomogeneous mixing of the elements in the ejecta, or emission
 of two shells with different abundances. 
We fitted less accurately the spectrum of day 35, which appears more complex.
 The emission lines on day 35 remained at least as broad as in the
 Chandra observation of days 18 and 23, while harder flux progressively
 emerged and lines due to higher ionization stages gradually appeared.
 However, the spectrum became also more complex and
 seemed  to indicate several different components. The emission lines
 in Fig. 5  are the same as those of
 the permanent supersoft X-ray source Cal 87 (see Orio et al. 2004), but 
 the lines were much broader in U Sco.

As shown in the compared fluxed spectra in Fig. 3,
 on day 35  N VI became much weaker. Fig. 3 also shows 
that in N VI He-like triplet and in the O VII triplet
 at day 35 there is also a definite shift
in the relative resonance, forbidden and intercombination
 line strengths since the February observations. In the
 February spectra, the resonance line is dominant,
but on the day 35, this was still true only for
 the newly appeared Ne IX triplet, while the resonance lines
 of the O VII and N VI triplets are weak. For the oxygen triplet,
 the intercombination line is clearly the  strongest.

The line ratios  did not became typical of ``simple'' photoionization,
 but we suggest they are explained with photoionization in the presence
of an additional strong UV source, which
 causes the strong intercombination line
 and  weak forbidden line (see Behar et al. 2004).
 While the unperturbed value (low density, no UV additional source)
of the {\it f/i} ratio is $\simeq$4.4, a much lower ratio like in our case indicates
 a nearby source of UV flux transferring flux
 from the forbidden to the intercombination line (Gabriel \& Jordan
 1969).
 Since these lines are not eclipsed, we cannot conclude that they were 
produced close to the WD, so at day 35
 at least some of the emission lines were due to 
photoionization, with a strong, additional
 UV flux from a nearby region. We hypothesize that this UV flux originated in 
hot condensations of shocked ejecta.

We also identify iron lines that cannot be explained in the context of either
collisional ionization or photo-ionization,
 no matter how high the iron abundance.
 These features
are Fe XVII at 16.78 \AA \ (already marginally detected
 on day 23),
 Fe XVIII at 14.207 \AA, and Fe XIX at 14.667 \AA.
 We note that these lines are very typical of quiescent, accreting CV's and have been 
observed also in
 in quiescent novae (Mukai et al. 2003, Mukai \& Orio 2005).  Mason et al. (2012)
 attribute also some specific emission features in optical
 spectra of U Sco on the very same day,
 to additional emission from the accretion disk. We hypothesize that also
 these iron lines, {  never detected before in previous nova outbursts}, arise in an accretion flow.
 It is significant they were observed in U Sco, which we know to have resumed accretion
 early during the outburst, and not in other erupting novae.

Finally, we mention two unsolved problems.
  Going back to the Chandra spectrum of day 18 we see at least one unidentified lines at 26.2 \AA \ and
 possibly one or two at 26.7 and 26.9 \AA. These lines are measured
 with low S/N, but since they
were marginally detected also in RS Oph (see Ness et al. 2011b, Table 5),
 they are likely to be real.
 The second puzzle concerns the N VII Ly$\alpha$ line, which merges with the weaker N VI He $\gamma$ line.
 On day 35,
 this appears to be a broader and more structured complex than all others. It is
certainly much broader than all other lines,
 and we can only speculate that this line may be produced in several different layers of ejecta with
 different velocity.

\section{Conclusions}

 We concluded that the absorption features observed with {\sl Chandra} arose
 in the WD atmosphere, and the velocity
 dropped close to zero at the base
 of the WD atmosphere in the following days, so that  the absorption lines were hidden almost at the
 center of the broad emission features in the spectra taken with
the {\sl XMM-Newton} RGS on days 23 and 35. 
 Fig. 7 shows in fact how the apparent red emission wings of the N VII and N VI lines 
 receded towards zero velocity, as their profile gradually became symmetric (Fig. 8).
 The profiles of emission lines that do not have corresponding
 strong WD atmospheric absorption
 features, in stark contrast with the N VI and VII features shown in Fig. 7,
 were almost symmetric around the rest wavelength from the beginning.
 
From these observations, we inferred that 
 the nova wind almost completely ceased between day 18 and 23 after the optical maximum.
 No other recurrent or classical nova has been known to have such a short
 mass loss period, but also the short optical light curve decay time and the 
large velocity measured in optical spectra were extreme for U Sco. 

 We agree with previous interpretations
 of the X-tray eclipse observed on day 23 and 35. It must be due to a Thomson
 scattering corona, and we propose that the eclipses, although not occurring
during the exposures,
 already existed on day 18.  Because of the inclination angle
 of U Scorpii,  we did not observe the hot WD directly,
 but we mostly measured the flux from Thomson scattered radiation,
which conserves the WD spectral shape and features, but gives 
 a lower limit on the true WD luminosity.

 Atmospheric models indicate that the WD was already extremely hot on day 
 18 and reached almost  1 million K on day 35. This high temperature is expected
 only on WD with mass close to the Chandrasekhar limit,
 above 1.3 M$_\odot$.  

We find that collisional ionization explains
 the X-ray emission lines observed
 in the February 2010 spectra of days 18 and 23.
 This conclusion is unexpected for a nova without copious circumstellar
 material around the system. In RS Oph, the red giant wind has
 left so much dense material,  that violent shocks can be expected (O'Brien \& Lloyd 1994, 
 Nelson et al. 2008). In U Sco the emission line spectrum is
 in the very soft energy range, unlike in RS Oph. In any case,
 the only conclusions we can draw from the X-ray spectral diagnostics
 are that mass loss was {\it not} a smooth process. Faster material
 probably collided with slower one emitted just a few days earlier, and 
 the ejecta had condensations
 or clumps of dense material that emitted
 the accretion lines. This may indicate also non isotropic outflows,
 or even small clumps like in the old shell of T Pyxidis (Schaefer et al. 2010). 
 
The line ratios and the absence of radiative
 recombination continua rule out photo-ionization
at least in February,
 but it seems clear that mass loss in this nova was
 not a smooth, continuous phenomenon. A description of the nova in terms
 of a steady radiation wind (see e.g. Hachisu \& Kato 2010 and references therein) 
 probably 
 describes only slow novae accurately, while in cases of fast and frequent outbursts
 the physics of the mass outflow seems to become quite complicated. 

On day 35, after mass loss from the system had ended,
the emission lines spectrum became more complex.  We identified 
 different components, including collisional ionization, photoionization
 in presence of an additional UV source, and 
emission lines of iron, associated with the resumed accretion and probably not with 
 the ejecta.  We have not been able to  fit all the 
 spectral characteristics of the three
 spectra accurately, but we propose a sort of ``zero order model'',
making a first step in the direction of physical understanding.
 We examined and analysed in this paper 
 the complexity and non homogeneity of a nova shell and its evolution.   

 X-ray grating observations of novae are precious not only because they
 allow to probe the WD peak temperature, temperature evolution and
 effective gravity, but also because of what we can learn on the mass
 outflow itself. It seems that there is still much work ahead of us - both
 modelling and observing new outbursts - in order to describe the nova physics accurately. 
 Novae are such interesting laboratories of extreme physics, that we think
 they are worth the effort.

\label{lastpage}

\end{document}